\documentclass[journal, hidelinks]{IEEEtran}
\usepackage[nocompress]{cite}
\usepackage{graphicx}
\usepackage[T1]{fontenc}
\usepackage[tt=false, type1=true]{libertine}
\usepackage[varqu]{zi4}
\usepackage[libertine]{newtxmath}

\usepackage{amsmath}
\usepackage{comment}
\usepackage{tabularx}
\usepackage{multirow, multicol}
\newcolumntype{Y}{>{\centering\arraybackslash}X} 
\usepackage{tabularray}
\usepackage{makecell}
\usepackage{array}
\usepackage{xspace}
\usepackage[font=small]{caption}
\usepackage{orcidlink}
\usepackage{nicefrac}
\usepackage{siunitx}
\usepackage{float}
\usepackage{enumitem}
\usepackage{subcaption}
\usepackage{relsize}
\usepackage{adjustbox}
\usepackage{soul}
\usepackage{amsthm}
\usepackage{mathtools}
\usepackage{etoolbox}
\usepackage{titlesec}
\usepackage{nicefrac}
\usepackage{diagbox}
\usepackage[hang, flushmargin]{footmisc}
\usepackage{algorithm}
\usepackage[noend]{algpseudocode}

\makeatother

\def\Def{\vcentcolon=}

\def\sothat{\,\lvert\,}
\newcommand\abs[1]{\lvert#1\rvert}\newcommand{\concat}{\ensuremath{\,+\!+\,}}

\newcommand{\kk}{\text{\scriptsize\sf k}\!\:}
\newcommand{\MM}{\text{\scriptsize\sf M}\!\:}
\newcommand{\hh}{\text{\scriptsize\sf h}\!\:}
\DeclareMathOperator*{\mean}{mean}

\def\hedge{\mbox{h\!\!\;-\!\!\;edge}\xspace}
\def\hgraph{\mbox{h\!\!\;-\!\!\;graph}\xspace}
\def\hedges{\mbox{h\!\!\;-\!\!\;edges}\xspace}
\def\hgraphs{\mbox{h\!\!\;-\!\!\;graphs}\xspace}

\newskip\paragraphbreak
\titleformat{\subsubsection}[runin]
  {\itshape} 
  {\thesubsubsection.}{0.5em}{}[: \quad] 

\titlespacing*\section{0pt}{*2}{*1}
\titlespacing*\subsection{0pt}{*2}{*1}
\titlespacing{\subsubsection}{0pt}{4pt}{0pt} 

\expandafter\def\expandafter\normalsize\expandafter{%
    \normalsize
    \setlength\abovedisplayskip{5pt} 
    \setlength\belowdisplayskip{5pt} 
    \setlength\abovedisplayshortskip{3pt} 
    \setlength\belowdisplayshortskip{3pt} 
}

\newlength{\itemizepadding}
\setlist[itemize]{topsep=\itemizepadding, leftmargin=3em} 
\setlist[enumerate]{topsep=\itemizepadding, leftmargin=3em} 

\renewcommand{\IEEEbibitemsep}{0pt plus 0.5pt}
\makeatletter
\renewcommand{\@biblabel}[1]{[#1]\hfill} 
\IEEEtriggercmd{\reset@font\normalfont\fontsize{8pt}{9pt}\selectfont} 
\IEEEtriggeratref{1}
\makeatother

\title{A Case for Hypergraphs to Model and Map{\linebreak}SNNs on Neuromorphic Hardware}

\author{
    Marco Ronzani \orcidlink{0009-0002-8485-0717}\,, Cristina Silvano \orcidlink{0000-0003-1668-0883}~\IEEEmembership{Fellow,~IEEE}
    \\DEIB, Politecnico di Milano, Italy\vspace{-18pt}
    \thanks{
        Manuscript received DD Month YYYY; revised DD Month YYYY; accepted DD Month YYYY. Date of publication DD Month YYYY; date of current version DD Month YYYY. (Corresponding author: Marco Ronzani.)
    }
    \thanks{
        The authors are with the Department of Electronics, Information and Bioengineering, Politecnico di Milano, Via Giuseppe Ponzio 34, 20133 Milano, Italy (e-mail: \url{marco.ronzani@polimi.it}; \url{cristina.silvano@polimi.it}).
    }
}

\begin{document}

\maketitle

\begin{abstract}
    Executing Spiking Neural Networks (SNNs) on neuromorphic hardware poses the problem of mapping neurons to cores.
    SNNs operate by propagating spikes between neurons that form a graph through synapses.
    Neuromorphic hardware mimics them through a network-on-chip, transmitting spikes, and a mesh of cores, each managing several neurons.
    Its operational cost is tied to spike movement and active cores.
    A mapping comprises two tasks: partitioning the SNN's graph to fit inside cores and placement of each partition on the hardware mesh.
    Both are NP-hard problems, and as SNNs and hardware scale towards billions of neurons, they become increasingly difficult to tackle effectively.
    In this work, we propose to raise the abstraction of SNNs from graphs to hypergraphs, redesigning mapping techniques accordingly.
    The resulting model faithfully captures the replication of spikes inside cores by exposing the notion of hyperedge co-membership between neurons.
    We further show that the overlap and locality of hyperedges strongly correlate with high-quality mappings, making these properties instrumental in devising mapping algorithms.
    By exploiting them directly, grouping neurons through shared hyperedges, communication traffic and hardware resource usage can be reduced beyond what just contracting individual connections attains.
    To substantiate this insight, we consider several partitioning and placement algorithms, some newly devised, others adapted from literature, and compare them over progressively larger and bio-plausible SNNs.
    Our results show that hypergraph-based techniques can achieve better mappings than the state-of-the-art at several execution time regimes.
    Based on these observations, we identify a promising selection of algorithms to achieve effective mappings at any scale.%
\end{abstract}

\begin{IEEEkeywords}
Spiking neural network, neuromorphic computing, HW mapping, hypergraph, partitioning, placement.
\end{IEEEkeywords}

\section{Introduction}\label{sec:intro}

Our biological brain is remarkably energy efficient in spite of its outstanding capabilities.
In light of this observation, \textbf{Spiking Neural Networks (SNNs)} have long since been under development to tap into the same event-driven type of intelligence.
At the same time, \textbf{NeuroMorphic Hardware (NMH)}, inspired by the asynchronous, sparse, and distributed nature of biological neurons, has been built to run SNNs with as little as tens of milliwatts per million neurons \cite{TrueNorthDesign}.

SNNs are composed of neurons connected to one another through synapses.
Each neuron integrates over time the spikes -- weighted signals -- it receives from its synapses and may eventually emit a spike of its own.
The network is usually operated in discrete time steps, where all incoming spikes are propagated and processed, while produced spikes are delivered to their target neurons and queued up for the next time step \cite{SNNHardwareImplementations, SNNSurvey}.
In practice, SNNs can be seen as a directed graph, with neurons for nodes and synapses as weighted edges.
The present work, in particular, raises such abstraction to that of a \textbf{directed hypergraph} by bundling together edges with a common source.

Most NMH is built with a 2D mesh of cores operating in parallel.
Each core handles multiple neurons, receiving spikes destined to them, updating their internal state, and emitting the spikes they produce.
Cores are interconnected via a network on chip that takes care of spike forwarding.
Example implementations include TrueNorth \cite{TrueNorthDesign}, Loihi \cite{Loihi}, SpiNNaker \cite{SpiNNaker}, and Neurogrid \cite{Neurogrid}.
For a SNN to run on a given NMH, each neuron must be assigned to a specific core by means of a \textbf{mapping}.
Such a mapping must carefully group connected neurons while respecting hardware limitations, like a maximum of neurons and synapses per core.

The main cost of running a SNN derives from the transmission of spikes between cores \cite{TrueNorthDesign}.
Each hop a spike takes consumes energy and adds latency between successive time steps of the system, that, to retain accuracy, must wait for all spikes to be delivered.
Nevertheless, when multiple neurons that are slated to receive the same spike are co-located in the same core, only a single copy of it is sent, relying on replication at the core level to save on communication.
Therefore, an optimal mapping is crucial to maximize the efficiency and throughput of the system.
A mapping is produced in two steps: \textbf{partitioning} of the SNN's graph, so that each partition can fit onto a virtual NMH core.
And \textbf{placement}, where each partition is assigned to a specific hardware core \cite{MappingVeryLargeSNNtoNHW}.
Both of these problems are known to be NP-hard \cite{FMpartitioning}.
Thus, with present SNNs exceeding millions of neurons, truly optimal solutions are out of reach.
Still, these problems have already been widely studied, under different constraints, for their role in electronic design automation and boolean function manipulation \cite{HypergraphPartAndClusteringSurvey, hMETIS_vlsi}, leading to plenty of \textbf{heuristics} to handle them effectively.
Hence, the expertise needed to handle them has long been established and available.

Several SNN mapping tools, integrating a variety of graph partitioning and placement heuristics, have been developed to seek near-optimal mappings.
In TrueNorth's ecosystem \cite{TrueNorthEcosystem}, the SNN is assumed to be partitioned at training time, and placement is done sequentially while minimizing the Manhattan distance between connected cores.
The Loihi compiler \cite{IntelMapper} tackles a variant of the problem that includes on-chip SNN training, filling cores successively by mapping nodes in order of connection strength with the present core.
Each core's cluster is seeded by the remaining node with the most inbound connections.
SpiNeMap \cite{SpiNeMap}, followed by DFSynthesizer \cite{DFSynthesizer}, employ the Kernighan-Lin recursive graph partitioning algorithm and a particle-swarm-based placement heuristic.
More recently, \cite{MappingVeryLargeSNNtoNHW} puts emphasis on scalability and layered SNNs, mapping billions of neurons through simple heuristics.
Partitioning proceeds sequentially with the nodes' natural order, saturating each partition before opening the next, while placement initially follows the Hilbert space-filling curve before force-directed refinement.
Similarly, EdgeMap \cite{EdgeMap} sequentially places each node in the partition currently minimizing its cut connections, then relying on a genetic algorithm for placement.
Finally, \cite{HierarchicalSplitMapping} experiments with hierarchical graph partitioning and synapse pruning, while \cite{StreamingGraphPartitioning} with streaming graph partitioners.

Nonetheless, the problem is far from solved.
On one side, many existing techniques have quadratic complexity in the number of SNN neurons or synapses, thus lacking scalability \cite{MappingVeryLargeSNNtoNHW}.
Others, instead, heavily rely on assumptions about the SNNs' topology to readily leverage the locality in their connections, lacking generalizability to non-layered networks, like liquid state machines \cite{LiquidStateMachine, SNNHardwareImplementations}.
In particular, we find that all existing tools model SNNs as graphs, rather than hypergraphs.
This leads them to exploit foremost the direct connections between neurons as a means to decide their proximity in the final mapping.
However, that only brings spike sources closer to their destinations, overlooking the origin and benefits of spike replication in NMH; namely, that it happens when multiple common destinations are in the same core.
What's more, when different destination neurons are brought together, they can share the hardware resources that handle inbound spikes -- something that does not occur between source and destination.
We also observe that reliance on individual edges leads some performance models to overestimate spike traffic altogether \cite{MappingVeryLargeSNNtoNHW, EdgeMap}.

For these reasons, we advocate a shift to the hypergraph model: it retains all the expressive power of graphs while explicitly capturing sets of nodes that partake in the same hyperedges and thus receive the same spikes, thereby exposing opportunities for replication and resource sharing.

In this work, we reformulate the problem of mapping spiking neural networks on neuromorphic hardware in terms of hypergraphs, focusing on fully capturing the circulation of spikes within their abstraction.
From this formalization, we identify two key properties of a good mapping, core-level synaptic reuse of spikes and locality of connected cores, and trace back their origin to two affinity measures defined between co-located nodes of a hypergraph.
Our intuition is that pursuing these properties serves as a guiding principle for effective mapping algorithms, making them not just descriptive of good solutions but prescriptive for their design.
Moreover, we empirically verify that good solutions indeed correlate with high exploitation of these two properties.

To act on our case for hypergraphs, we actively leverage them in several mapping algorithms that apply our intuition.
Currently, much of the literature on hypergraphs remains to be explored when applied to the SNN-on-NMH mapping problem \cite{HypergraphPartAndClusteringSurvey, KaHyPar, HypergraphsSNNinHPCPhDThesis, HypergraphsSpectralTheory}.
Drawing from it, we implement a hierarchical partitioning scheme adapted to the constraints of NMH and introduce a placement scheme based on a discretization of the hypergraph's spectrum, which naturally closes the gap between highly connected partitions.
We also revisit within our framework several partitioning and placement techniques from existing SNNs mapping tools: sequential partitioning, Hilbert curve, force-directed, and minimum distance placement.
In addition, we propose a new heuristic for hypergraph partitioning that, driven by the synaptic reuse property, works by bundling together nodes that share membership in many hyperedges.
For each algorithm, we study its computational complexity and trade-offs in terms of speed versus quality of results.
Finally, we conduct experiments to evaluate all combinations of the presented heuristics and the role hypergraph properties played in each of them.
In doing so, and to the best of our knowledge, we are the first to include recurrent and biologically plausible SNNs, alongside classic feedforward ones, in our validation.
This enables us to identify the most suitable set of algorithms to handle the mapping problem for current and future large-scale SNNs.

\vspace{\paragraphbreak}
\noindent
This paper is organized as follows: in Section~\ref{sec:intro}, we provide the context and motivation for our work.
In Section~\ref{sec:preliminaries}, we introduce our notation for SNNs and NMH, while in Section~\ref{sec:mapping_model}, we formalize the mapping problem accordingly.
With Section~\ref{sec:algorithms}, we present existing and novel techniques to handle \hgraph partitioning and placement.
Then, in \mbox{Section~\ref{sec:experimental_evaluation}}, we discuss our experimental evaluation and results.
Finally, with Section~\ref{sec:conclusion}, we summarize and comment on our findings.

\section{Preliminaries}\label{sec:preliminaries}

\begin{figure*}[t]
    \centering
    \vspace{-4pt}
    \includegraphics[width=\textwidth]{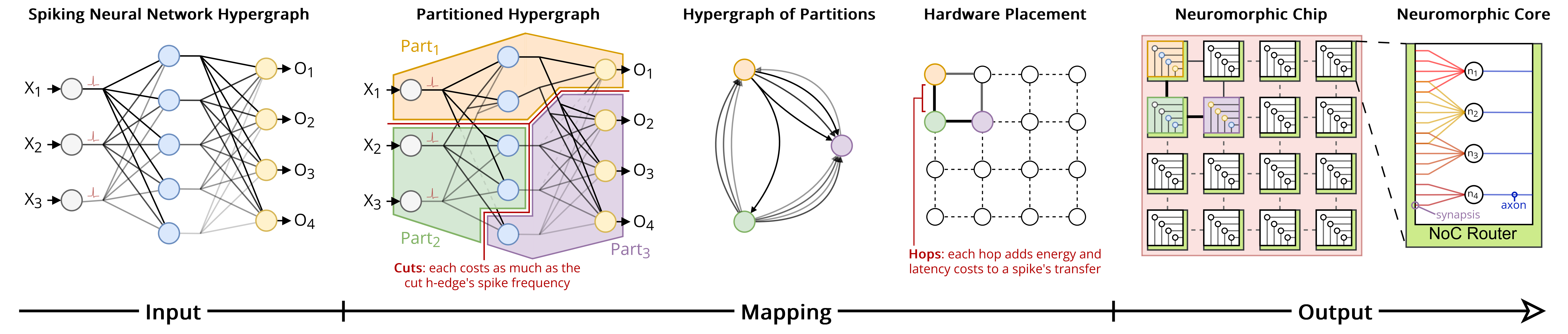}
    \vspace{-14pt}
    \caption{Mapping steps: from a spiking neural network to its partitioning and placement on neuromorphic hardware.}
    \vspace{-14pt} 
    \label{fig:mapping_steps}
\end{figure*}

\subsection{Spiking Neural Networks}\label{subsec:snns}


Spiking neural networks are a computational paradigm for artificial intelligence inspired by the structure and functions of biological neurons.
Unlike Artificial Neural Networks (ANNs), they retain and leverage the asynchronous and distributed calculation properties found in nature.
As such, they propagate information through events, represented by spikes, that leave a neuron through its axon to then be received by other neurons connected to it via weighted synapses.
Upon receiving a spike, a neuron updates its internal mechanisms and may emit another spike of its own.
For simplicity, spike propagation and generation of future spikes are typically performed in discrete time steps \cite{SNNSurvey}.

SNNs can be built in several ways.
They can be converted from a feedforward ANN, retaining most of the original model's accuracy \cite{SNNtoolbox, SpikingJelly}.
Crucially, this results in layered SNNs, with distinct, ordered groups of neurons corresponding to the original network's layers and all synapses concentrated in between those groups \cite{SNNHardwareImplementations, MappingVeryLargeSNNtoNHW}.
Alternatively, SNNs can be trained from scratch.
While methods in this field have yet to produce models with accuracy rivaling ANN on traditional datasets, they consistently find success in event-driven and temporal tasks \cite{SNNvsANN, SNNSurvey}.
In those natively trained SNNs, we may find any arbitrary topology of synapses, consider, for example, liquid state machines, which present cycles in the form of feedback connections \cite{LiquidStateMachine, FeedbackSNNs, OnlineLSMMapping}.
Such bio-like \mbox{SNNs have been called cyclic, feedback, or recurrent}.

We represent a SNN as a directed weighted hypergraph (\textit{\hgraph}), where each hyperedge (\textit{\hedge}) has a single source.
Thus, nodes represent neurons, while \hedges the axons departing once from each neuron.
\begin{equation}\label{eq:snn_graph}
    \begin{gathered}
        G_S \Def (N, E_S, w_S) \text{,} \\
        E_S \Def \{e = (s, D) \sothat s \in N, D \subseteq N\} \text{,} \\
        w_S : E_S \rightarrow \mathbb{R}^+ \text{.}
    \end{gathered}
\end{equation}
Let $G_S$ be the \hgraph, each $n \;\!\!\in\;\!\! N$ a node and each $(s, D) \;\!\!\in\;\!\! E_S$ one of its \hedges.
Then, $s$ is the \hedge's source while $D$ is the set of its destinations.
Lastly, $w_S$ is a function assigning a weight to each \hedge, corresponding to its spike frequency in the SNN.
From each neuron of a SNN departs a single axon; thus, any $s \in N$ will appear exactly once as a source among \hedges.
This means there is~a~\mbox{one-to-one} correspondence between \hedges and neurons, and an \hedge's spike frequency can also be interpreted as that of its source neuron.

\subsection{Neuromorphic Hardware}\label{subsec:neuromorphic_hardware}


\begin{figure}[b]
    \centering
    \includegraphics[width=0.65\columnwidth]{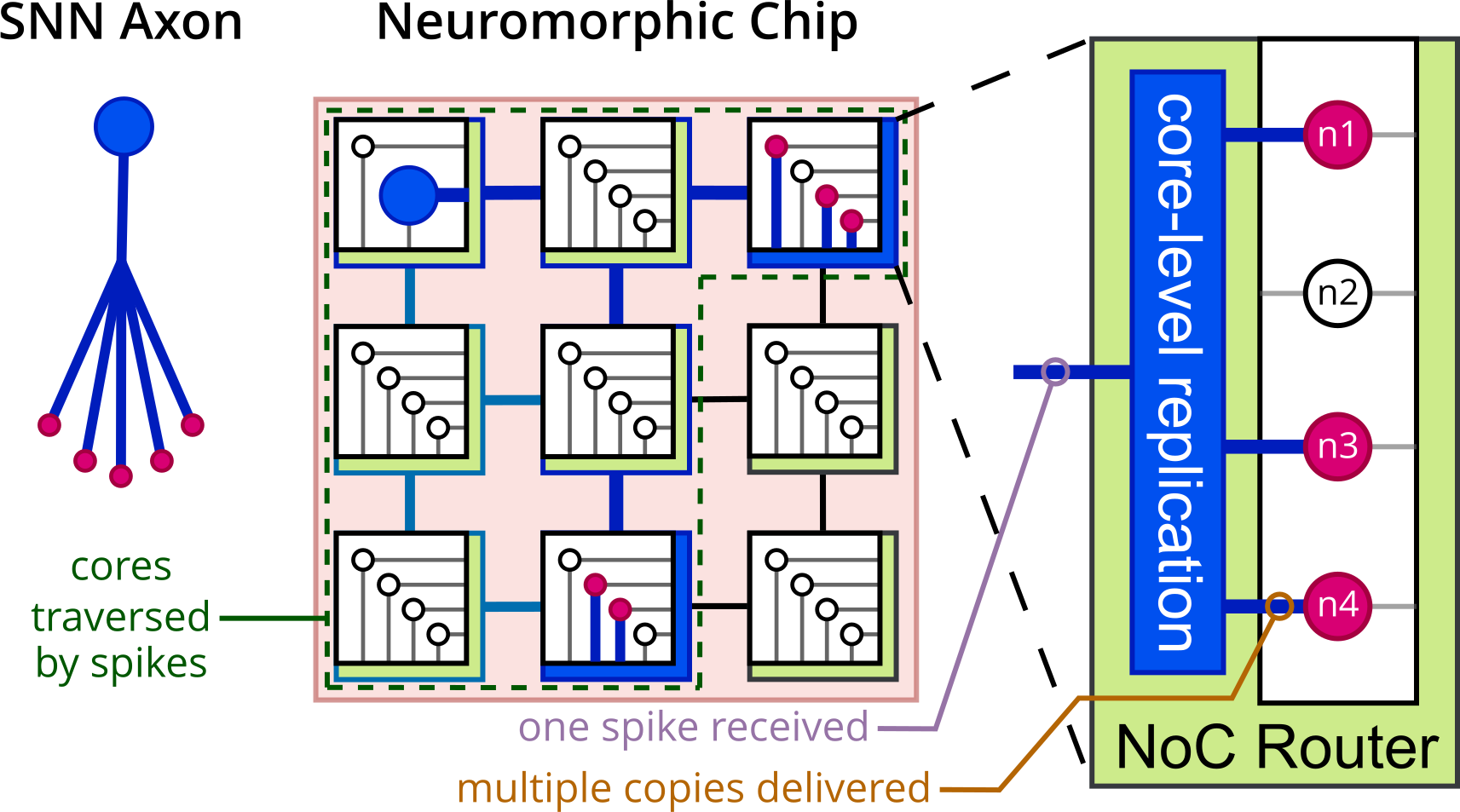}
    \caption{Example of axon mapping and core-level spike replication.}
    \vspace{-9pt}
    \label{fig:hardware_replication}
\end{figure}

Neuromorphic hardware accelerators for SNNs consist of a spatial mesh of cores. 
A core handles several neurons, storing their parameters, updating them upon spike arrival, and sending out produced spikes.
Every core has a router that connects it to the others through a network on chip and takes care of spike multicast and forwarding between different cores.
Global synchronization only occurs across the computation's discrete time steps, once all cores have handled the spikes generated by the previous step.
Otherwise, to minimize power consumption, NMH mimics SNNs in being mostly asynchronous, with a core activating only upon receiving a spike.
Consequently, its major performance bottlenecks are the energy and time spent on spike movements, spike transmission latency in particular determines the global time step duration and thus the system's throughput \cite{TrueNorthDesign, SNNHardwareImplementations, IntelMapper}.
To mitigate these costs, NMH cores support spike replication: each core receives at most one instance of any spike and internally dispatches it to as many neurons as needed.
This limits the copies of a spike that may need to circulate to one per core.
See Fig.~\ref{fig:hardware_replication} for an example of such multiple local deliveries.
Such architecture is typically scalable to multiple chips connected in a higher-order mesh.
However, since off-chip communication is less predictable than that on-chip, this work's hardware model focuses on single-chip systems.

Formally, NMH is here modeled as a 2D lattice of cores of known width and height:
\begin{equation}\label{eq:hardware}
    H \Def \{h = (x, y) \in \mathbb{N}^2 \sothat 0 \leq x < width, 0 \leq y < height\} \text{.}
\end{equation}
Each core $h \in H$, represented by its coordinates, can handle a maximum of $C_{npc}$ neurons and $C_{apc}$ incoming axons in common between such neurons.
In other terms, when mapping a SNN's \hgraph over the hardware, at most $C_{npc}$ nodes can share a common core, and their distinct inbound \hedges can't be more than $C_{apc}$.
Each source of spikes seen by a core usually needs a dedicated hardware queue, resulting in the above limits on the maximum number of axons that neurons in a core can collectively connect to.
Then, depending on the implementation, inside a core, each neuron could be allowed to have a synapse with all such axons, as in TrueNorth \cite{TrueNorthDesign}, or there might be a further limit on the number of synapses per core, $C_{spc} < C_{npc} \cdot C_{apc}$, as in Loihi \cite{Loihi}.

\section{Mapping Model}\label{sec:mapping_model}

Given a SNN $G_S$, we model one of its possible partitionings through a surjective function $\rho : N \rightarrow P$ that maps each node of the SNN to a partition $p \in P$.
By pushing forward $G_S$ through $\rho$, we define $G_P$ as the new \hgraph that emerges among the partitions of $G_S$:
\begin{equation}\label{eq:part_graph}
    \begin{gathered}
        G_P \Def (P, E_P, w_P) \text{,} \\
        P \Def \{\rho(n) \sothat n \in N\} \text{,} \\
        E_P \Def \{e_P = (\rho(s), \{\rho(d) \sothat d \in D\}) \sothat e_S = (s, D) \in E_S\} \text{,} \\
        w_P(e_P) \Def w_S(e_S) \text{.}
    \end{gathered}
\end{equation}
In the so-obtained \hgraph $G_P$, each node $p \in P$ is a partition assigned by $\rho$, hence $\abs{P} \leq \abs{N}$.
Then, \hedges between partitions are constructed as $E_P$ and, for each \hedge, its weight is defined by $w_P : E_P \rightarrow \mathbb{R}^+$.
We may subsequently merge \hedges with identical source and destinations by adding together their weights.
Here, the one-to-one relation between \hgraph nodes and \hedges is lost, as multiple \hedges may depart from the same partition.
A partitioning is valid iff it fits the hardware's constraints.
The preimage size of each codomain element of $\rho$, being the size of each partition, is limited by $C_{npc}$:
\begin{equation}\label{eq:npc_constraint}
    \forall \, p \in P \;\, \abs{\{n \in N \sothat \rho(n) = p\}} \leq C_{npc} \text{.}
\end{equation}
The count of distinct \hedges inbound to the same partition is limited by $C_{apc}$:
\begin{equation}\label{eq:apc_constraint}
    \forall \, p \in P \;\, \abs{\{(s, D) \in E_S \sothat \exists \, n \in D, \rho(n) = p\}} \leq C_{apc} \text{.}
\end{equation}
The overall count of inbound connections to the same partition is bounded by $C_{spc}$:
\begin{equation}\label{eq:spc_constraint}
    \forall \, p \in P \;\, \textstyle\sum_{(s, D) \in E_S} \abs{\{d \in D \sothat \rho(d) = p \}} \leq C_{spc} \text{.}
\end{equation}

The cost function to minimize in an optimal partitioning is the \textit{weighted connectivity}, or $\lambda-1$ metric, between partitions:
\begin{equation}\label{eq:connectivity}
    Conn(G_P) = \sum_{e_P = (s, D) \in E_P} w_P(e_P) \cdot \abs{D} \text{.}
\end{equation}
This equates to minimizing the number of partitions each \hedge connects, by paying once the \hedge's weight per connected partition \cite{KaHyPar}.
Incidentally, this is the same as minimizing the average amount of spikes that need to transit between cores once the SNN is mapped on NHW \cite{MappingVeryLargeSNNtoNHW}.

\begin{table}[b]
    \centering
    \vspace{4pt}
    \resizebox{\columnwidth}{!}{
        \begin{tblr}{colspec={|X[c, m]|X[5.2, m]|}, row{1,7} = {c}, row{8} = {rowsep = 0pt}, row{9, 10} = {rowsep = 4pt}, width=1.1\columnwidth}
            \hline
            \textbf{Symbol} & \textbf{Definition} \\
            \hline
            $\| \cdot \|$ & $L1$ norm (Manhattan distance) \\
            \hline
            $E_R, L_R$ & Energy and latency for a spike's routing \\
            \hline
            $E_T, L_T$ & Energy and latency for a spike's transmission between two cores \\
            \hline
            $\tau(h, h_s, h_d)$ & Probability of a spike being routed through core $h$ when going from core $h_s$ to $h_d$ (see \cite{MappingVeryLargeSNNtoNHW} for details) \\
            \hline
            $Rect(h_1, h_2)$ & Set of lattice points contained in the closed square defined by the opposite corners $h_1$ and $h_2$ \\
            \hline
            \hline
            \textbf{Metric} & \textbf{Expression} \\
            \hline
            \makecell{Energy \\ consumption} & \raisebox{3pt}{$\displaystyle \sum_{e = (s, D) \in E_P} \: \sum_{d \in D} w_P(e) (\lVert \gamma(s) - \gamma(d) \rVert (E_R + E_T) + E_R)$} \\
            \hline
            \makecell{Average \\ latency} & $\displaystyle \frac{\sum_{e = (s, D) \in E_P} \sum_{d \in D} w_P(e) (\lVert \gamma(s) - \gamma(d) \rVert(L_R + L_T) + L_R)}{\sum_{e \in E_P} w_P(e)}$ \\
            \hline
            \makecell[t]{Average \\ congestion} & \raisebox{-1pt}{$\displaystyle \sum_{e = (s, D) \in E_P} \: \sum_{d \in D} \: w_P(e) \sum_{h \in Rect(\gamma(s), \gamma(d))} \tau(h, \gamma(s), \gamma(d))$} \\
            \hline
        \end{tblr}
    }
    \vspace{-2pt}
    \caption{SNN mapping performance metrics. Adapted from \cite{MappingVeryLargeSNNtoNHW}.}
    \vspace{-6pt}
    \label{tab:costs}
\end{table}

\begin{table}[b]
    \centering
    \begin{minipage}[t]{0.39\columnwidth}
        \centering
        \resizebox{\linewidth}{!}{
            \begin{tblr}{colspec={|X[0.5, c, m]|X[0.5, c, m]l|}, column{3} = {wd = 0pt, leftsep = 0pt, rightsep = 0pt}, rowsep = 1pt} 
                \hline
                \SetCell[r=2]{c} \textbf{NMH~Cost} & \SetCell[r=2]{c} \textbf{Value} & \phantom{\textbf{Value}} \\
                \cline{2-3}
                & & \phantom{small} \\
                \hline
                $E_R$ & 1.7 pJ & \phantom{$width$,~$height$} \\
                \hline
                $L_R$ & 2.1 ns & \phantom{$C_{npc}$} \\
                \hline
                $E_T$ & 3.5 pJ & \phantom{$C_{apc}$} \\
                \hline
                $L_T$ & 5.3 ns & \phantom{$C_{spc}$} \\
                \hline
            \end{tblr}
        }
        \vspace{-2pt}
    \end{minipage}%
    \hspace{0.01\columnwidth}
    \begin{minipage}[t]{0.59\columnwidth}
        \centering
        \resizebox{\linewidth}{!}{
            \begin{tblr}{colspec={|X[c, m]|X[0.5, c, m]|X[0.5, c, m]|}, rowsep = 1pt}
                \hline
                \SetCell[r=2]{c} \textbf{Constraint} & \SetCell[c=2]{c} \textbf{Value} & \\
                \cline{1-3}
                & small & large \\
                \hline
                $C_{npc}$ & 1024 & 4096 \\
                \hline
                $C_{apc}$ & 4096 & 65536 \\
                \hline
                $C_{spc}$ & 16384 & 262144 \\
                \hline
                $width$,~$height$ & \SetCell[c=2]{c} 64$\,\times\,$64 & \\
                \hline
            \end{tblr}
        }
        \vspace{-2pt}
    \end{minipage}
    \caption{Reference hardware costs and constraints \cite{MappingVeryLargeSNNtoNHW, Loihi}.}
    \vspace{-4pt}
    \label{tab:hw_costs}
\end{table}

Finally, a placement (layout) is modeled by another injective function $\gamma : P \rightarrow H$, assigning (placing) each partition to a hardware core.
Fig.~\ref{fig:mapping_steps} summarizes all discussed steps.
The post-layout performance metrics minimized by an optimal mapping are listed in Table~\ref{tab:costs}.
They are an adaptation to hypergraphs of the formulae in \cite{MappingVeryLargeSNNtoNHW} and include the system's energy consumption, latency, and interconnect congestion.
Their estimation is conditioned on low-level measurements of the hardware's routers and wires that build up the cost of carrying spikes between cores.
We take the measurements used in our experiments, as well as the hardware constraints, from Intel's Loihi \cite{Loihi} ("small") and \cite{MappingVeryLargeSNNtoNHW} ("large"), see Table~\ref{tab:hw_costs}.

The present mapping model ensures that spike replication capabilities of NMH cores are accounted for.
This occurs when multiple neurons inside the same partition (and eventually, core) share an \hedge.
In such a case, the hardware only transmits the \hedge's spikes once towards the core, relying on it to, very cheaply, integrate the spike on each neuron needing it.
For instance, Eq.~\ref{eq:connectivity} accounts for this by paying an \hedge's weight in connectivity only once per partition.
Note that the cost of input spikes has not been modeled, as those only arrive once in hundreds or thousands of system steps, thus having a negligible impact on the final metrics \cite{SNNHardwareImplementations}.



\subsection{Mapping Properties}\label{subsec:mapping_properties}

\begin{figure}[t]
    \centering
    \includegraphics[width=\columnwidth]{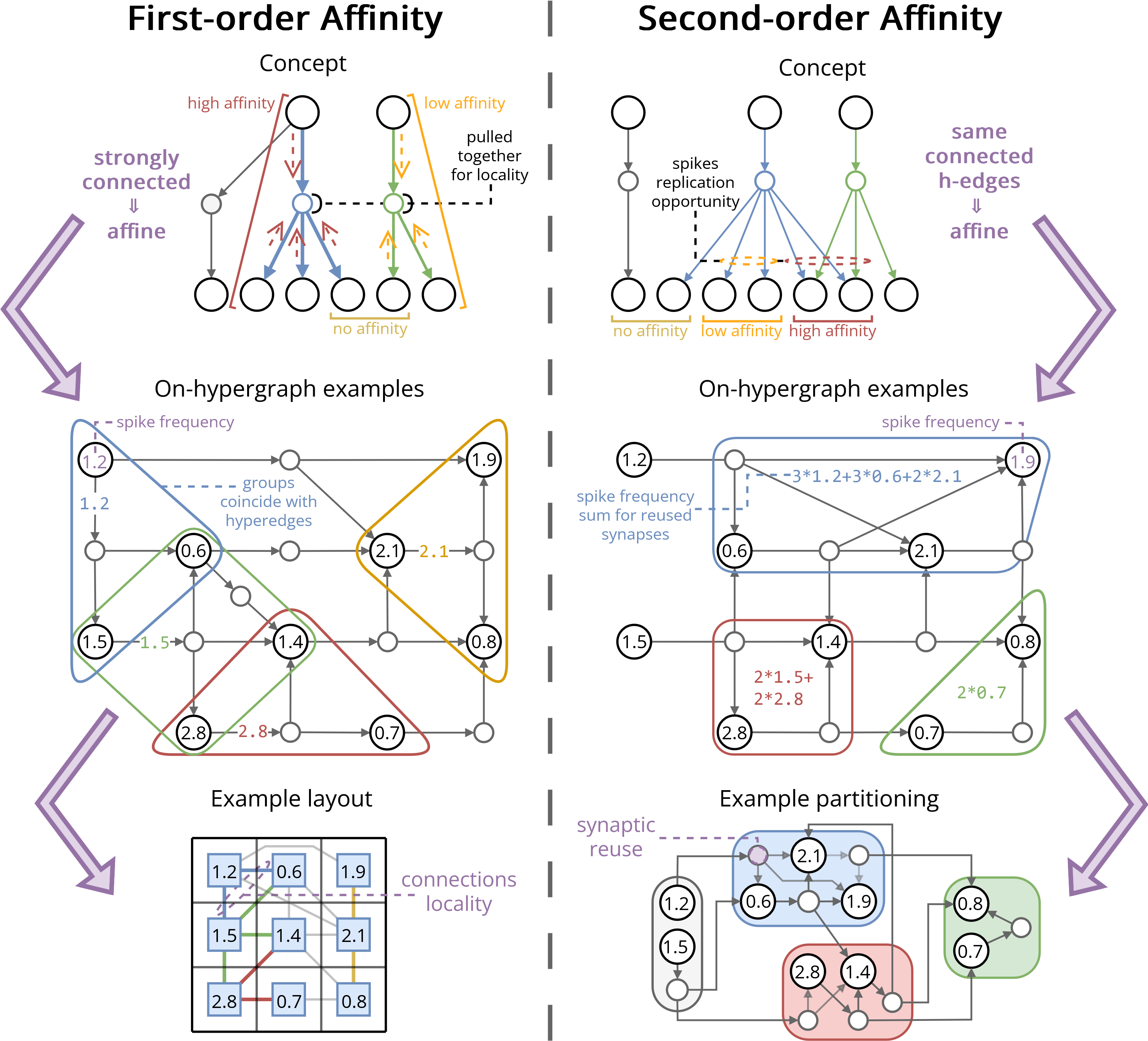}
    \vspace{-12pt}
    \caption{Illustration and examples of first- and second-order affinity.}
    \label{fig:affinity}
\end{figure}

From the presented formalization and objectives, we identify two properties that characterize good mappings.
A mapping's partitioning shall minimize connectivity by maximizing the replication of spikes inside hardware cores, inherently reducing the number of copies of a spike that are transmitted.
This manifests as \textit{synapses reusing spikes} -- a property we call \textit{synaptic reuse} -- when multiple neurons in the same partition, and thus hardware core, receive spikes from a common source (axon).
In turn, this implies that the core only allocates physical resources to receive such a source's spikes once, hence being able to fit more nodes.
A mapping's layout shall then minimize the above metrics by limiting the overall distance traveled by inter-core spikes, thereby reducing traffic on the interconnect.
This leads to the property of \textit{connections~locality}, reflecting how closely the destinations of an \hedge are placed -- ideally within a small neighborhood of cores.
The more confined an \hedge is, the less congestion its spikes cause, leading to more consistent, timely deliveries and reduced latency across global time-steps.
This also seamlessly creates an opportunity for the hierarchical multicasting of spikes, on architectures that implement such a feature \cite{Loihi}.
Connections locality is the layout counterpart of synaptic reuse, that relies on co-member nodes getting closer, rather than sharing the same core.
Ultimately, synaptic reuse and connections locality are complementary, both being fundamental mechanisms to drive mapping algorithms.

At the \hgraph level, we thus pinpoint two measures that can effectively guide the construction of mappings.
These come in the form of affinities that advocate for certain nodes to be mapped in close proximity of one another.
\textit{First-order affinity}, the weight of the direct connection between all pairs of nodes under the same \hedge.
The stronger the connection, the closer such nodes shall be to reduce its cost.
\textit{Second-order affinity}, the weight of co-membership, that is, the cumulative weight of the \hedges a set of nodes partakes in together \cite{LINE}.
In other words, grouping nodes with overlapping sets of connections amortizes their \hedges' connections cost over them.
Synaptic reuse emerges when high-second-order-affinity nodes share the same partition.
Conversely, connections locality is reliant on the relative adjacency of nodes with high first-order affinity.
As a result, second-order affinity shall guide the partitioning, while first-order affinity the placement.
Fig.~\ref{fig:affinity} provides a visual intuition for this.
Finally, the key challenge of heuristics is identifying which affine nodes it is best to map closer together while performing as few iterations as possible over the \hgraph.

\vspace{\paragraphbreak}
\noindent
In summary.
To reduce \textbf{cuts} during \textbf{partitioning}, a mapping exploits spike \textbf{replication} in NMH cores through \textbf{synaptic reuse} by grouping nodes with high \textbf{second-order affinity}.
To limit spike \textbf{hops} after \textbf{layout}, via shorter \textbf{Manhattan distances} on NMH, a mapping increases \textbf{connections locality} by pulling closer partitions with high \textbf{first-order affinity}.


\subsection{Towards Hypergraphs}\label{subsec:towards_hypergraphs}

In graph form, a SNN’s source node connects to multiple destinations with a different edge for each, thus losing access to the notions of first- and second-order affinity.
Practically, this hinders the quantification and realization of synaptic reuse and connections locality, where you need to reason in terms of overlapping connections and distance between co-member nodes.
With \hgraphs, instead, \hedges naturally group together nodes under a joint source, exposing these properties.
In fact, porting to \hgraphs the metrics in Eq.~\ref{eq:connectivity} and Tab.~\ref{tab:costs} from \cite{MappingVeryLargeSNNtoNHW} corrected an overestimation of $w_P$ that did not account for synaptic reuse due to the elicitation of edges, rather than \hedges.
Ultimately, the graph model's lack of a native coalesced representation of destinations with a shared source adds overhead to algorithms that rely on some measure of affinities for guidance.
Partitioning struggles to identify nodes with overlapping inbound sets, while placement cannot pull together \hedges as a whole and uniformly close the gap between all their nodes.
In turn, this encourages mappings to solely rely on source-destination connections, missing out on reuse and locality.
Arguably, building an index that groups edges by source could mitigate the issue.
However, we object that doing so already transitions from a graph towards the proposed hypergraph abstraction.
In this sense, many approaches already subtly lean towards hypergraphs, but, by not fully adopting them yet, they encounter limitations in problem modeling and algorithm design.

For these reasons, we concluded that hypergraphs are better suited to model SNNs throughout the mapping process.
Nonetheless, it is trivial to decompose each \hedge in multiple edges to use a graph representation if needed. 
As a last note, the hypergraph formulation more closely mirrors biological neural networks, with each \hedge corresponding to an axon.

\section{Proposed Algorithms}\label{sec:algorithms}

\begin{figure*}[t]
    \centering
    \vspace{-4pt}
    \includegraphics[width=\textwidth]{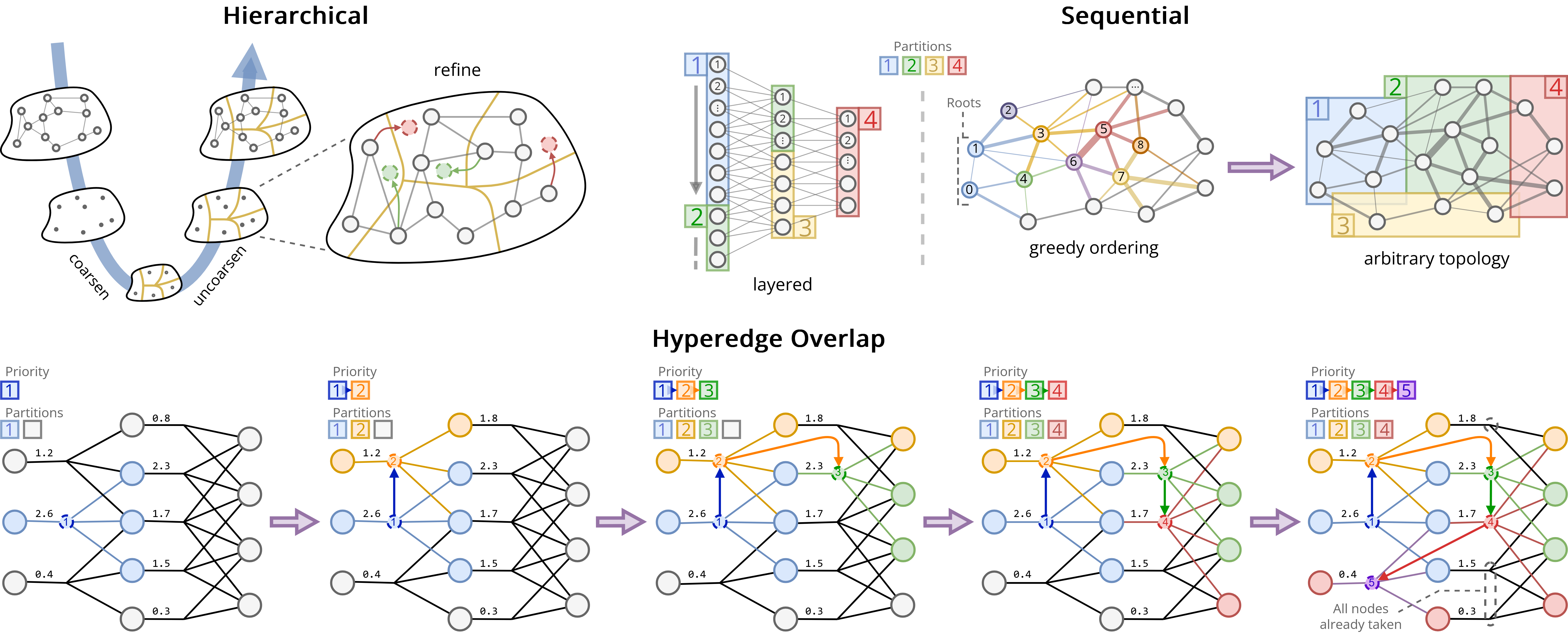}
    \vspace{-12pt} 
    \caption{Overview of the considered partitioning algorithm.}
    \vspace{-14pt}
    \label{fig:partitioning}
\end{figure*}



Both constrained \hgraph partitioning and placement are NP-hard problems; thus, it is computationally unfeasible to aim for an optimal solution when millions of nodes and \hedges are involved.
On account of this, SNN mapping techniques shall consist of heuristics that, paired with suitable assumptions, trade off execution time and solution quality \cite{ComputersAndIntractability}.
Even more so, heuristics must be scalable, retaining a feasible time-to-solution (e.g. tens of hours) on a single machine while striving for near-optimal mappings, even when growing SNNs to "human-scale" -- tens of billions of neurons.
As such is the end goal of neuromorphic computing \cite{TrueNorthDesign}.
We will now describe several heuristics.
Some are adapted from research on SNN mapping, others are based on hypergraphs or EDA studies that have not yet been applied in this context.
Moreover, we highlight how each heuristic makes effective use of the properties defined in Sec.~\ref{subsec:mapping_properties}.

Henceforth, we rely on a generic input \hgraph to ease the discussion of algorithmic complexity: let $n$ be its number of nodes, $e$ its number of \hedges, and $d$ the average \hedge cardinality.
From which, each node is on average touched by $h \Def \nicefrac{e \cdot d}{n}$ \hedges.
Actually, with our definition of SNNs (Sec.~\ref{subsec:snns}) we have $n = e$ and so $d = h$\;\!; we will use this fact to standardize all reported complexities.
In general, $n, e >> d, h$, as SNNs tend to favor localized connectivity \cite{SNNSurvey}, thus the complexity of examined algorithms will foremost be linear in the \hgraph's size, $n$ or $e$.
Consequently, it will be the additional dependencies on $d$, $h$, or other constants that make up the difference in final running time.

When outlining algorithms, we adopt the following data structures.
Let an \hgraph be represented as a list of \hedges, each \hedge being a pair consisting of a list of nodes, whose first element is the source, and a spike-frequency weight.
Nodes are identified by consecutive integers.
Lastly, two auxiliary indices provide constant-time access to the set of \hedges inbound to a node and to its outbound \hedge \cite{KaHyPar_origins}.



\subsection{HyperGraph Partitioning}\label{subsec:partitioning}


Hypergraph partitioning involves separating an \hgraph's nodes in disjoint sets while minimizing the number and cardinality of hyperedges that get split among multiple partitions.
This problem can come in many forms depending on its constraints and quality metrics.
In our case, we have upper limits on nodes and connections per partition, distinct inbound \hedges per partition, and total partitions, but we are allowed to arbitrarily reduce the number of partitions.
This "distinct" keyword, especially, is what enables~us~to~fit~many nodes per partition, saving cuts by leveraging synaptic reuse.
The ideas behind presented algorithms are shown in Fig.~\ref{fig:partitioning}.

\subsubsection{Hierarchical Partitioning}\label{subsubsec:hierarchical_partitioning}

Inspired by hMETIS \cite{hMETIS_k_way} and KaHiPar \cite{KaHyPar_origins, KaHyPar}, this is a multi-level hypergraph partitioning method with refinement based on the Fiduccia-Mattheyses algorithm \cite{FMpartitioning}.
Seeing that existing implementations aim to create an exact, given number of balanced partitions \cite{HypergraphPartAndClusteringSurvey}, we shall rework them to minimize partitions under the constraints from Sec.~\ref{sec:preliminaries}.
The method begins by constructing a hierarchy of progressively coarser representations of the input SNN hypergraph through weighted edge-coarsening \cite{hMETIS_vlsi}.
This leads to an initial partitioning that is then refined while the hierarchy is undone.
Crucially, such a multi-level approach makes the use of iterative refinement techniques, usually reserved for small \hgraphs, feasible.

During a coarsening round, nodes are iterated in random order.
For each node, all yet unpaired nodes partaking in the same \hedges are considered and scored by the total weight of the \hedges linking them.
Essentially, a pair-wise evaluation of second-order affinity.
A pair is formed between the present node and, if any, the highest scoring one such that the two together do not violate per-partition constraints.
Before the next round, each pair is placed in the same partition and subsequently treated as a single node, while duplicate \hedges are merged.
Coarsening stops when no pairs can be formed or there are exactly $\lceil \nicefrac{\abs{n}}{C_{npc}} \rceil$ nodes, providing the initial, almost minimized number of partitions; let it be $k$.

Each node of the final coarsened graph is a partition of our candidate solution.
Such partitioning is then incrementally projected back onto finer levels by replacing each node with the pair that formed it, if any.
During each uncoarsening round, local refinement is applied using a gain-based heuristic.
The algorithm evaluates whether moving a neuron to a neighboring partition (e.g. one its outbound \hedges connect to) reduces the overall communication cost as defined by Eq.~\ref{eq:connectivity}.
Nodes are visited in random order, and all advantageous moves are greedily applied so long as constraints are respected.
Still, the refinement process is efficient, as coarse moves implicitly act on multiple nodes, while cost variations remain computed locally.

The result is an $O(e \cdot d^2 \:+\: e \cdot d \cdot k)$ complexity, dominated by coarsening: $O(n \cdot h \cdot d) = O(e \cdot d^2)$.
Refinement, instead, runs in $O(n \cdot h \cdot k \:+\: e \cdot d) = O(e \cdot d \cdot k)$, achieved by precomputing node counts per partition per \hedge via a single scan of all \hedges, costing $O(e \cdot d)$.
The hierarchy spans $O(log\,\nicefrac{n}{k})$ levels, each roughly halving the \hgraph's size, amounting to a small multiplicative constant to the complexity.
Notably, there exist approaches that avoid enumerating all node pairs during coarsening, removing the quadratic dependency on $d$ by approximating pair affinity \cite{KaHyPar}.
However, we opted for the original, exact, edge-coarsening for two reasons: first, our implementation strictly relies on coarsening to form a minimal number of partitions for the given constraints, which is different from standard literature where $k$ is fixed a priori.
Second, in this way, our work can effectively depend on hierarchical partitioning as a source of near-optimal solutions for more scalable algorithms to compare against.

\subsubsection{Hyperedge Overlap-based Partitioning}\label{subsubsec:hyperedge_overlap_partitioning}


This is our novel greedy algorithm that constructs one partition after the other, placing nodes in them while solely focusing on second-order affinity to maximize synaptic reuse.
This is achieved through a single sweep of all \hedges, for each laying all its nodes.
To address reuse, the order of \hedges is dynamically adjusted such that the next one always has the highest overlapping set of connected nodes with those seen immediately before it.
Alg.~\ref{alg:hyperedge_overlap_partitioning} presents the full approach\footnotemark[1].

\footnotetext[1]{
    Convention: $argmax^*$\:mean\:any\:arbitrary\:maximizer,\;likewise\:for\:$argmin^*$. Then, $argmin^{lex}$ does the same but on tuples ordered lexicographically.
}

First, we sort \hedges in descending order of their number of connections (line~\ref{alg:hop:line:sort}).
This will be the order in which our outer loop sees \hedges, thus prioritizing those that contribute most to connectivity by involving many nodes.
We also set up an initial empty partition (line~\ref{alg:hop:line:initial_partition}), where we will allocate nodes until further doing so would violate constraints, causing the creation of a new partition (line~\ref{alg:hop:line:new_partition}).
Then, for each \hedge, we go over its destination nodes and assign them to the current partition.
This is done by continuously selecting the node whose set of inbound \hedges introduces the fewest new inbound \hedges to the current partition; in case of a tie, the node with the largest inbound set is chosen, thereby maximizing overlap after minimizing additions (line~\ref{alg:hop:line:node_selection}).
Exceptionally, we first also assign the \hedge's source node, if it isn't part of any other \hedge (line~\ref{alg:hop:line:include_input_nodes}).
Simultaneously, an addressable priority queue is used to track the number of times each yet-unseen \hedge appears among those connected to the assigned nodes.
In particular, the queue tracks, for each still unvisited \hedge, the ratio between the number of nodes assigned to the current partition that it touches and the \hedge's count of connected, but still unassigned, nodes (line~\ref{alg:hop:line:priority_queue_update}).
Such ratio is then scaled by the \hedge's spike frequency to become the priority (line~\ref{alg:hop:line:pop_queue}).
When the priority queue is not empty, it diverts the outer loop's behavior, making the next \hedge to be handled the one with the highest priority (line~\ref{alg:hop:line:next_hedge}).
Upon instantiation of a new partition, the queue is flushed (line~\ref{alg:hop:line:flush_queue}).
Execution terminates after having visited all \hedges.

\begin{algorithm}[t]
    \smaller
    \caption{Partitioning by Hyperedge Overlap}\label{alg:hyperedge_overlap_partitioning}
    \begin{algorithmic}[1] 
        \Require $G_S = (N, E_S, w_S)$, $C_{npc}$, $C_{apc}$, $C_{spc}$, $\abs{H}$
        \Ensure $\rho : N \rightarrow P$ 
        \For{$n$ \textbf{in} $N$}
            \State $\rho(n)$ = $p_{null}$ \Comment{blank partitions assignment}
            \State inbound$(n)$ = $\{(s, D) \in E_S \sothat n \in D \}$ \Comment{precompute inbound sets}
            \State outbound$(n)$ = $\{(s, D) \in E_S \sothat s = n\}$ \Comment{outbound singletons}
        \EndFor
        \For{$e$ \textbf{in} $E_S$}
            \State size$(e = (s, D))$ = $\abs{D} + 1$ \Comment{remaining\:connections\:count\:per\:\hedge} 
            \State pq$(e)$ = $0$ \Comment{addressable priority queue}
        \EndFor
        \State sort($E_S$, key : size($\cdot$), descending) \label{alg:hop:line:sort}
        \State seen = \{ \} \Comment{already visited \hedges}
        \State $p_{curr}$ = next-partition$(P)$ \label{alg:hop:line:initial_partition}
        \State npc = 0, spc = 0, apc = \{ \} \Comment{current partition's constrained variables} 
        \While{$\abs{\text{seen}}$ < $\abs{E_S}$}
            \If{$\exists \, a \in E_S$ s.t. pq$(a)$ > $0$} \label{alg:hop:line:next_hedge}
                \State $e$ = $(s, D)$ = $argmax^*_{b \in E_S \setminus \text{seen}} \, w_S(b) \cdot \text{pq}(b)$ \Comment{pop from queue} \label{alg:hop:line:pop_queue}
            \Else
                \State $e$ = $(s, D)$ \textbf{first in} $E_s \setminus \text{seen}$ \Comment{pop from sorted $E_S$}
            \EndIf
            \State seen = seen $\cup$ $\{e\}$
            \State nodes = $\{d \in D \sothat \rho(d) = p_{null}\}$ \Comment{unassigned destinations only}
            \If{inbound$(s)$ = $\varnothing$}
                nodes = nodes \!\!\;$\cup$\!\!\, $\{s\}$ \Comment{include\:input\:nodes} \label{alg:hop:line:include_input_nodes}
            \EndIf
            \While{$\abs{\text{nodes}} > 0$}
                \State $n$ = $argmin_{m \in \text{nodes}}^{\text{lex}} (\abs{\text{inbound}(m) \setminus \text{apc}}, -\abs{\text{inbound}(m)})$ \label{alg:hop:line:node_selection} 
                \If{npc = $C_{npc}$ \textbf{or} spc > $C_{spc}$ \textbf{or} $\abs{\text{apc} \cup \text{inbound}(n)}$ > $C_{apc}$}
                    \State \textbf{assert} npc > 0 \Comment{else neuron $n$ can't fit core constraints} 
                    \State pq$(a)$ = $0$ $\forall \, a \in E_S$ \Comment{clear the queue} \label{alg:hop:line:flush_queue}
                    \State $p_{curr}$ = next-partition$(P)$ \label{alg:hop:line:new_partition}
                    \State npc = 0, spc = 0, apc = \{ \}
                    \State \textbf{continue}
                \EndIf
                \State npc += 1, spc += 1, apc = apc $\cup$ inbound$(n)$
                \State $\rho(n)$ = $p_{curr}$
                \State nodes = nodes $\setminus$ $\{n\}$
                \For{$c$ \textbf{in} $\text{(inbound}(n) \cup \text{outbound}(n)) \setminus \text{seen}$} \Comment{visit \hedges} \label{alg:hop:line:visit_hedges}
                    \State pq$(c)$ = (pq$(c)$*size$(c) + 1$)/(size$(c) - 1$) \Comment{occurrences / size} \label{alg:hop:line:priority_queue_update}
                    \State size$(c)$ -= 1
                \EndFor
            \EndWhile
        \EndWhile
        \State \textbf{assert} $\abs{P}$ < $\abs{H}$ \Comment{else exceeded the partition count limit} 
        \State \textbf{return} $\rho$
    \end{algorithmic}
\end{algorithm}

Here, the queue's priority is an incremental proxy measure of second-order affinity; as such, it focuses on the spike-frequency-weighted fraction of co-membership exhibited in each remaining \hedge by the nodes in the current partition.
Crucially, by this definition, the priority can be computed with only the limited extra burden of iterating once over each node's connections.
The fallback on \hedges ordered by size, instead, exists to rapidly fill the queue while assigning as many nodes as possible early on, speeding up subsequent iterations.
Then, our node selection policy within each \hedge strictly ensures maximum synaptic reuse while prioritizing a snug fit inside constraints; for this reason, it overlooks the spike frequency.
This results in overall fuller partitions with high spike replication, while spike frequency returns to prominence when the next \hedge is extracted from the priority queue.
In general, visiting an \hedge assigns only its destination nodes, since doing so for the source would require all its inbound \hedges to follow it in the current partition.
But depending on the topology, a source might have a very different inbound set, easily saturating the partition, in turn winding up apart from its affine nodes.
However, nodes that receive outside inputs don't present any inbound \hedge; as such, we are free to place them close to their destinations.

Even if this approach mainly iterates over \hedges, that is but a means to visit and assign nodes in an order that favors affinity.
In fact, each node is processed, and its connections visited, only once (line~\ref{alg:hop:line:visit_hedges}), giving a complexity of $O(n \cdot h \cdot log(C_{npc} \cdot h))$.
Where the logarithm is the cost of managing the heap implementing the priority queue, that, being reset with each partition, can contain at most $h$ \hedges per neuron in the partition.
Thus, by hiding the relatively small constant $C_{npc}$, expanding the definition of $h$, and using $n = e$, we can rewrite the complexity as $O(e \cdot d \cdot log\,d)$.

\subsubsection{Sequential Partitioning}\label{subsubsec:sequential_partitioning}

\begin{algorithm}[t]
    \smaller
    \caption{Greedy Nodes Ordering}\label{alg:greedy_ordering}
    \begin{algorithmic}[1]
        \Require $G = (N, E, w)$ \Comment{here $G$ can be either $G_S$ or $G_P$}
        \Ensure ordered$(N)$
        \State $L$ = (\:) \Comment{empty ordered nodes list}
        \For{$n$ \textbf{in} $N$}
            \State pq$(n)$ = $0$ \Comment{addressable priority queue}
            \State inbound$(n)$ = $\{(s, D) \in E \sothat n \in D \}$ \Comment{precompute inbound sets}
            \State outbound$(n)$ = $\{(s, D) \in E \sothat s = n\}$ \Comment{outbound singletons}
        \EndFor
        \For{$n$ \textbf{in} $argmin_{m \in N}(\abs{\text{inbound}(m)})$}
            \State pq$(n)$ = $+\infty$ \Comment{start from min-inbound-set nodes}
        \EndFor
        \While{$\abs{L} < \abs{N}$}
            \If{$\exists \, m \in N \setminus L$ s.t. pq$(m)$ > $0$}
                \State $n$ = $argmax^*_{m \in N \setminus L}$ pq$(m)$ \Comment{pop from queue}
            \Else
                \State $n$ = $argmin^*_{m \in N}(\abs{\text{inbound(m)}})$ \Comment{default:\:min\!\!\;-\!\!\;inbound\!\!\;-\!\!\;set\:node}
            \EndIf
            \State $L$ = $L \concat \, n$
            \For{$e = (s, D)$ \textbf{in} outbound$(n)$ \textbf{for} $m$ \textbf{in} $D$}
                \State pq$(m)$ += $w(e)$
            \EndFor
        \EndWhile
        \State \textbf{return} $L$
    \end{algorithmic}
\end{algorithm}

A fast strategy from \cite{MappingVeryLargeSNNtoNHW} that requires nodes to be ordered, it then puts successive nodes in the same partition as long as NMH constraints allow for it, and only upon violating them does it initiate a new partition.
Clearly, this becomes effective only if nodes can be ordered such that successive ones have almost identical connection patterns with the rest of the \hgraph.
The nodes' overlapped connections, in turn, would lead to synaptic reuse and hardware resource sharing, requiring fewer partitions and lowering the partitioning's connectivity cost.

Obtaining such a convenient ordering is, however, just as hard as solving the partitioning problem.
Therefore, we adopt two strategies.
For ANN-derived SNNs, nodes are constructively ordered by iterating through the ANN's layers and sequencing the neurons inside each layer.
Hence, the resulting order will naturally present similar inbound and outbound \hedges among neighboring nodes \cite{MappingVeryLargeSNNtoNHW}.
For example, consider the overlap between the receptive fields of two neighboring output neurons in a convolution.
While, for arbitrary SNNs, no such order is intrinsically available.
Ergo, it needs to be constructed, but finding an optimal one that clusters nodes with highly overlapping inbound connectivity -- maximizing synaptic reuse between nearby nodes in the ordering -- reduces to a variant of the minimum linear arrangement problem, which is itself NP-hard~\cite{ComputersAndIntractability}.
Consequently, we settle for a greedy approximation: we keep an addressable priority queue of nodes, initially filled with those having the least inbound \hedges (usually those with none), all with the highest priority.
We then iteratively build the order by removing the queue’s top element and adding all the nodes it is connected to to the queue, each with priority equal to the spike frequency of its connection.
If a node is already present in the queue, its priority is increased by such spike frequency.
Refer to Alg.~\ref{alg:greedy_ordering} for details\footnotemark[1].
This approach yields an ordering with high local synaptic reuse by repeatedly grabbing the next most connected node to those already seen, exactly what sequential partitioning expects to exploit.
Its complexity is $O(n\,log\,n \cdot h) = O(e \cdot d \cdot log\,n)$; the logarithm emerges from implementing the queue as a heap.


In the end, sequential partitioning is most effective when the input SNN has a layered-like structure, but it can be efficiently generalized to any SNN.
In particular, once the ordering is available, this is the fastest partitioning technique we hereby consider, with complexity $O(n)$.

\subsection{Initial Placement}\label{subsec:initial_placement}

The placement determines which SNN partition is assigned to which NMH core.
Since the interconnect of NMH propagates spikes along rows and columns, the objective is to minimize the Manhattan distance between heavily connected partitions.
Heavily connected partitions striving to get closer means that their common \hedges will tend to remain confined within a small neighborhood of cores, which is the aforementioned concept of connections locality.

\begin{figure}[t]
    \centering
    \vspace{-2pt}
    \includegraphics[width=\columnwidth]{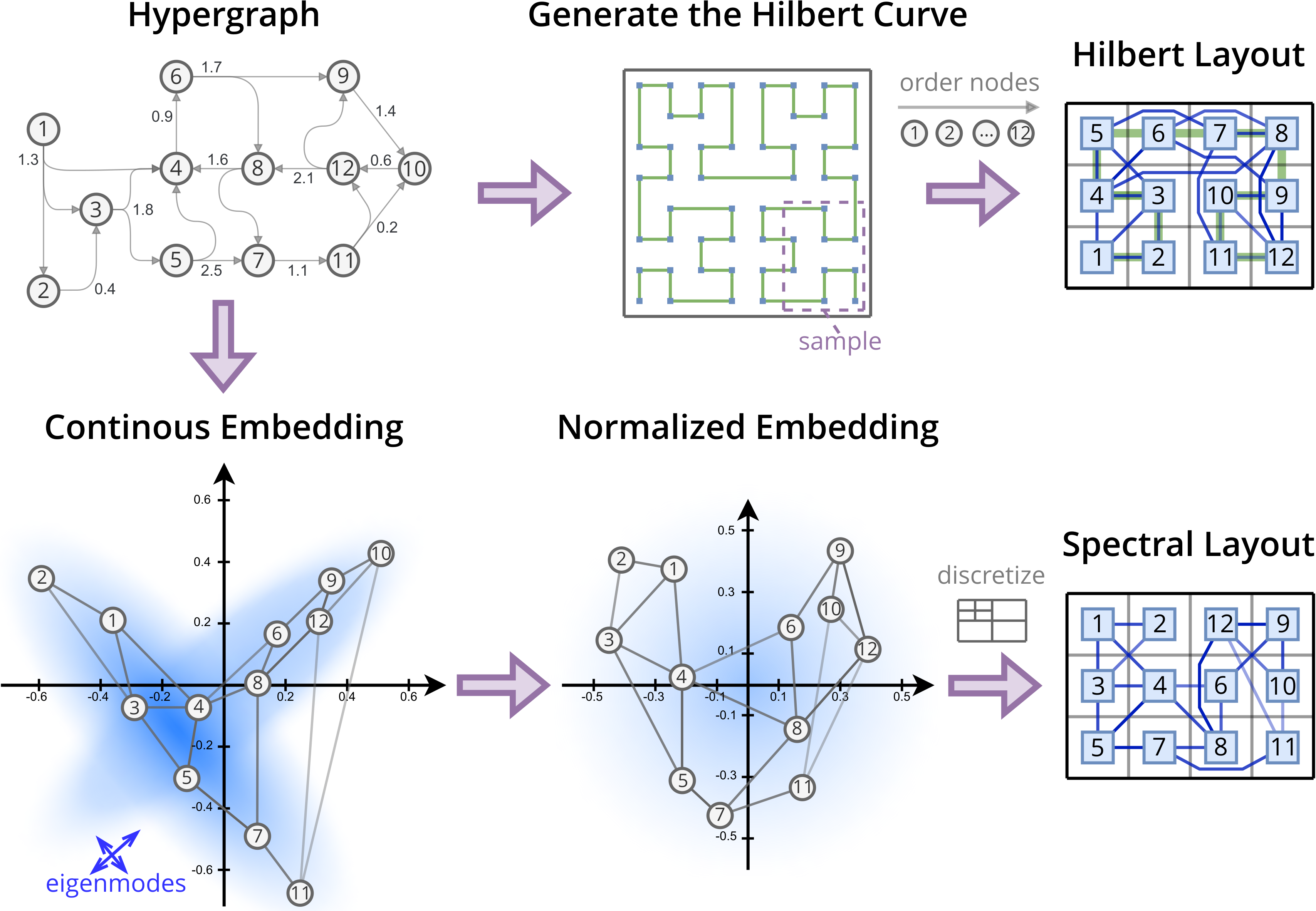}
    \vspace{-12pt} 
    \caption{Example constructions of Hilbert and spectral placements.}
    \label{fig:initial_layout}
\end{figure}

The purpose of an initial placement is to provide a good starting point for placement heuristics by attempting to maximize the locality of cores that share many \hedges.
Not all placement routines can leverage an initial placement, but when that is possible, providing a good one is crucial.
To this end, we consider two algorithms that preserve locality while laying \hgraph nodes on a lattice.
First, the discrete Hilbert Space-filling Curve \cite{MappingVeryLargeSNNtoNHW}.
Second, the use of spectral embeddings to project the \hgraph structure into 2D Euclidean space.
An example of both is given in Fig.~\ref{fig:initial_layout}.

\subsubsection{Hilbert Space-filling Curve}\label{subsubsec:hilbert_initial_placement}

The discrete Hilbert space-filling curve provides a mapping from a 1D sequence to 2D integer coordinates.
Moreover, it does so while preserving the sequence's locality, as in neighboring elements in the sequence are mapped to points in close spatial proximity \cite{MappingVeryLargeSNNtoNHW}.
Assuming to have a sequence of nodes with high connections locality, this gives us an easy transition from 1D to 2D, resulting in a good initial definition for $\gamma$.
But of course, to preserve locality, we need to have locality in the first place.

Unlike the original SNN, its partitioned \hgraph is likely to present cycles, and we can't know a priori if its nodes will present some natural order with strong locality.
Therefore, we employ the same ordering techniques discussed in Sec.~\ref{subsubsec:sequential_partitioning} for arbitrary SNNs.
When the partitioned \hgraph is acyclic, we construct its topological ordering using a queue-based variant of Kahn’s algorithm \cite{KahnAlgorithm}, where roots are enqueued first and, at each step, outgoing \hedges are processed in decreasing weight order before newly freed nodes are added to the queue.
This is typically the case of layered SNNs that underwent sequential partitioning, as in \cite{MappingVeryLargeSNNtoNHW}.
Otherwise, our greedy order from Alg.~\ref{alg:greedy_ordering} is used.
Since constructing the Hilbert curve has linear cost in the number of coordinates spanned, this method's final complexity is $O(e \cdot d)$ for acyclic graphs, else $O(e \cdot d \cdot log\,n)$, as in Sec.~\ref{subsubsec:sequential_partitioning}.

\subsubsection{Spectral Placement}\label{subsubsec:spectral_initial_placement}

To compute an initial compact and structure-aware placement of an \hgraph onto a discrete 2D lattice, we can rely on the \hgraph's spectrum, in the form of the eigenvalues and eigenvectors of its Laplacian matrix.
The idea is to first project the \hgraph's connectivity geometry into a two-dimensional coordinate system, using the lowest nontrivial eigenmodes of the Laplacian as a guide.
Then, we just need to scale and discretize such an embedding over our lattice.
The so-obtained layout captures both first- and second-order affinity: the Laplacian-induced quadratic smoothness objective simultaneously causes strongly connected nodes to collapse together and penalizes the variance within each \hedge \cite{GraphSpectralDrawing, HypergraphsSpectralTheory}.
Concretely, we construct the normalized Laplacian of the hypergraph $\mathcal{L} \in \mathbb{R}^{\abs{P} \times \abs{P}}$ \cite{GraphSpectralDrawing} by exploding each \hedge in individual connections:
\begin{equation}\label{eq:laplacian}
    \begin{gathered}
        \mathcal{L}_{i,j} \Def
        \begin{cases}
            \phantom{+} 1 & \text{if } i = j \\
            - \dfrac{1}{\sqrt{wdeg(i)\, wdeg(j)}} \sum\limits_{\substack{e = (s, D) \in E_P \\ \{i,j\} \subseteq \{s\} \cup D}} w_P(e) & \text{if } i \neq j
        \end{cases} \\[6pt]
        \text{where } wdeg(p) = \sum_{e = (s, D) \in E_P \sothat p \in \{s\} \cup D} w_P(e) \text{.}
    \end{gathered}
\end{equation}
From it, we compute the two eigenvectors corresponding to the smallest non-zero eigenvalues \cite{LaplacianEigenmaps}:
\begin{equation}\label{eq:eigenvectors}
    (\mathbf{u}_1, \mathbf{u}_2) \;\;\;\text{s.t.}\;\;\; \mathcal{L}\mathbf{u}_k = \lambda_k \mathbf{u}_k, \quad 0 < \lambda_1 \leq \lambda_2 \leq \cdots \text{,}
\end{equation}
The resulting eigenmodes provide the smoothest nontrivial orthogonal directions minimizing the quadratic objective:
\begin{equation}\label{eq:smoothness}
    \mathcal{Q}(\Gamma) \;\Def\; \text{tr}(\,\Gamma^\top \mathcal{L} \Gamma\,) \text{ where } \Gamma = \left[ \gamma_c(p)^\top \right]_{p \in P} \,\text{,}
\end{equation}
which encodes a penalty proportional to the dispersion within each \hedge~-- not just pairwise links, but whole \hedge spread -- and, as such, it precisely boosts connections locality.
These two eigenvectors ($\,\Gamma$ columns) each have one entry per node and together define a 2D coordinate for every node, one component from each eigenvector ($\,\Gamma$ rows).
So:
\begin{equation}\label{eq:gamma_continous}
    \Gamma = \left[ \mathbf{u}_1, \mathbf{u}_2 \right] \text{, and thus } \gamma_c(p) = (\mathbf{u}_{1,p}, \mathbf{u}_{2,p})
\end{equation}
is our initial continuous embedding, with $\gamma_c : P \rightarrow \mathbb{R}^2$ being the continuous dual of $\gamma$.
This $\gamma_c$ is first normalized to fit within the unit square, then scaled to occupy a compact, nearly-square, rectangular region encompassing enough points of $H$ to fit all partitions.
The compact layout is centered within the grid, and each node’s continuous position is then discretized to the nearest unoccupied integer lattice point, at last yielding $\gamma$.
To ensure that each node is assigned a unique coordinate without collisions, a KD-tree is used to efficiently search for the nearest available grid point, and assigned points are removed from the candidate set as placement proceeds.
Nodes are visited in descending order of total spike frequency for their bound \hedges.

The final result is a $\gamma$ to grid coordinates that preserves the structural affinities of the partitioned \hgraph while minimizing the total edge length under Manhattan distance.
If the Laplacian is represented in sparse form, computing the $k$ smallest nonzero eigenpairs with an iterative method requires $O((n + e \cdot d) \cdot k)$ operations over a small number of iterations \cite{ARPACK}.
Here, $k = 2$, and $e \cdot d$ is the number of non-zeros in the Laplacian.
Then, normalization and scaling are $O(n)$, and the discretization step is $O(n\,log\,n)$ due to nearest-neighbor search with a KD-tree.
Overall: $O(e \cdot d \,+\, n\,log\,n)$.



\subsection{Placement Refinement}\label{subsec:placement_refinement}

\begin{figure}[t]
    \centering
    \vspace{-1pt}
    \includegraphics[width=0.98\columnwidth]{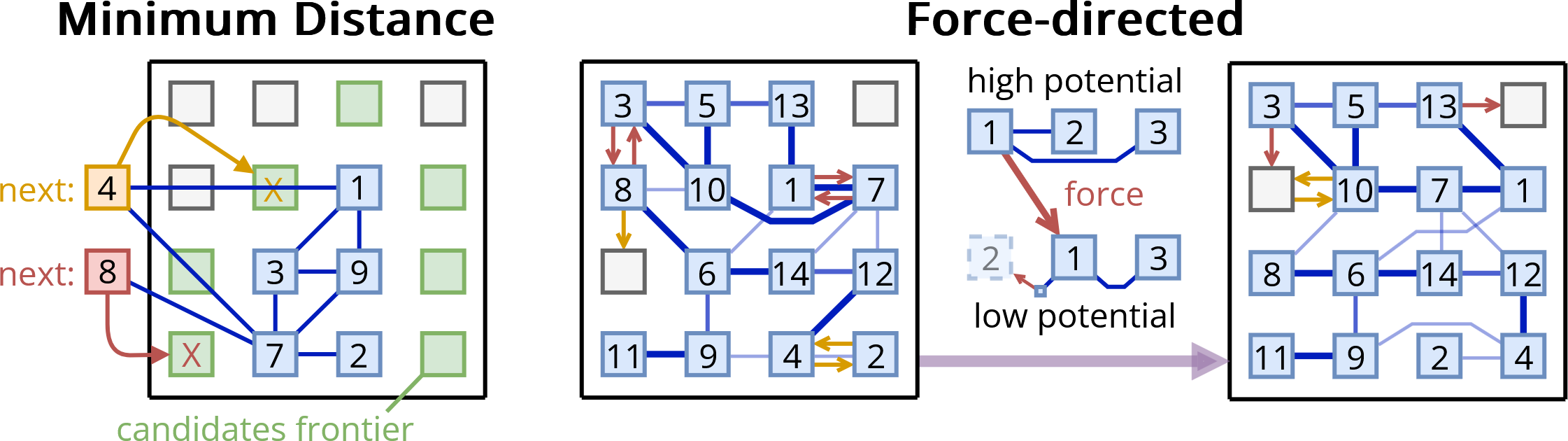}
    \vspace{-2pt}
    \caption{Overview of the considered placement refinement algorithms.}
    \vspace{-2pt}
    \label{fig:placement_refinement}
\end{figure}

With a good initial placement, costly refining algorithms can be efficiently applied on a local basis to improve it.
We take one such algorithm from literature \cite{MappingVeryLargeSNNtoNHW}.
Then, we also include an algorithm from \cite{TrueNorthEcosystem} that directly goes from \hgraph to placement, with no need for an initial solution.
Fig.~\ref{fig:placement_refinement} depicts the basic ideas of both algorithms.

\subsubsection{Force-Directed}\label{subsubsec:force_directed_placement}


Given an initial placement, the idea is to efficiently identify pairs of neighboring partitions such that if their placements were swapped, connections locality would improve.
In \cite{MappingVeryLargeSNNtoNHW} it is proven that minimizing, across all partitions, a virtual potential of $G_P$ and $\gamma$:
\begin{equation}\label{eq:potential}
    Pot_\gamma(p) \Def \sum_{e = (s, D) \in E_P \sothat p \in D} \| \gamma(p) - \gamma(s) \| \cdot w_P(e) \text{,}
\end{equation}
is equivalent to minimizing the total system energy and latency as defined in Table~\ref{tab:costs}.
This leads to defining a force as the delta in the potential when a partition is moved (not yet a swap, temporary overlaps are accepted at this point) by one core along one of the four cardinal directions:
\begin{equation}\label{eq:force}
    \begin{gathered}
        Force_{\gamma}(p, v) \Def Pot_\gamma(p) - Pot_{\gamma'}(p) \text{,} \\
        \text{where } \gamma'(q) =
        \begin{cases}
            \gamma(q) & \text{if } q \neq p \\
            \gamma(p) + v & \text{if } q = p
        \end{cases} \\ 
        \text{and } v \in \{(1, 0), (-1, 0), (0, 1), (0, 1)\} \text{.}
    \end{gathered}
\end{equation}
As a result, a placement can be optimized by swapping partitions between neighboring cores so long as the sum of opposing forces pulling on the two is positive.

The refinement routine, adapted from \cite{MappingVeryLargeSNNtoNHW}, simply iterates on all pairs of neighboring cores whose partitions give a positive sum of forces, swaps them, and repeats, until no such pairs remain.
Candidate pairs are visited by higher-force first.
Recomputing all forces after a change is costly; thus, they are lazily updated only on candidate partitions before swapping their placements and on all partitions connected to swapped ones.
We improve this technique by not only~attempting swaps between placed partitions, but between any unused core and surrounding ones with a placed partition as well.
In Eq.~\ref{eq:force}, this means that $\gamma(p) + v$ only needs to be valid -- inside $H$ -- and not necessarily in $\gamma$'s codomain already.
In practice, this change allows the placement's active cores to change when not all cores are utilized.
Then, we modify the evaluation of Eq.~\ref{eq:potential} to use $max(\| \cdot \|, 1)$ instead of $\| \cdot \|$, to ensure that when calculating deltas in the potential, any co-located partitions still present a unit Manhattan distance.
Otherwise, their force's effect would be neglected, potentially causing endless loops of positive-force swaps.

One iteration of the above routine has costs $O(n \cdot d \cdot h^2)$, since each force calculation is linear in $h$ and a swap can cause the update of up to $2 \cdot d \cdot h$ forces among other partitions.
Assuming $t$ iterations, the final complexity is $O(t \cdot n \cdot d \cdot h^2)$.
That said, $t$ is hard to estimate a priori, with our experiments seeing it vary between 50 and 1.5\kk.
Still, refinement can be terminated early if the achieved performance is deemed sufficient, exposing $t$ as a parameter.

\subsubsection{Minimum Distance Placement}\label{subsubsec:minimum_distance_placement}


This is the placement algorithm used in TrueNorth \cite{TrueNorthEcosystem}.
Input partitions, those receiving spikes from outside the system, are spread out as much as possible while remaining centered and evenly spaced between themselves and the cores' lattice's borders.
Then, each subsequent partition is placed on the core that minimizes its total Manhattan distance from all already-placed partitions it is connected to.
This is performed while going through partitions in topological order, ensuring that all sources of \hedges inbound to the current node have already been placed, thus making the raw Manhattan distance a viable proxy for the final mapping performance.
Once again, we use a queue-based variant of Kahn’s algorithm \cite{KahnAlgorithm} to obtain the topological order for acyclic \hgraphs and our greedy techniques presented in Alg.~\ref{alg:greedy_ordering} when the partitioned graph is cyclic.
Such a visit of nodes works since, for the Manhattan distance to approximate mapping performance, it is enough to see the highest spike frequency \hedges.

The way it was presented in \cite{TrueNorthEcosystem}, this method required that the SNN be trained, from scratch, as if already partitioned; we took the liberty to generalize it to arbitrary SNNs by pairing it with other partitioning algorithms.
Moreover, we found two ways to improve this algorithm over its original version.
First, we can weight each Manhattan distance by the total spike frequency of the \hedges connecting the two partitions.
Second, rather than searching all cores for the one of lowest total Manhattan distance, we look exclusively at those immediately adjacent to cores that are already in use.
This means that to place a partition, we just need to visit the frontier around $\gamma$'s growing codomain.

This is a very straightforward technique that has a compelling complexity of $O(n \cdot h \cdot \abs{H} + q)$.
Here, $n \cdot h$ is the cost of computing the distance from the present partition to all those it is connected to, that could all already be placed, and $q$ is either $O(e \cdot d)$ for acyclic graphs, else $O(e \cdot d \cdot log\,n)$, for the topological ordering.
Although restricting the next partition's candidate placements to the active core's frontier does not reduce the asymptotic complexity, it makes the contribution of $\abs{H}$ negligible in practice.
Combined with its linear dependence on $n$ and $h$, this makes the minimal distance placement the most scalable across our experiments.

\section{Experimental Evaluation}\label{sec:experimental_evaluation}

\subsection{Experimental Setup}\label{subsec:experimental_setup}

\begin{table}[t]
    \centering
    \resizebox{\columnwidth}{!}{
        \begin{tblr}{colspec={|X[0.1, m]|X[1.1, m]|X[0.5, c, m]X[1.0, c, m]X[1.0, c, m]X[0.95, c, m]|}, row{1} = {c}, width=1.1\columnwidth, rowsep = 1pt}
            \hline
            \SetCell[c=2]{c} \textbf{Network} & & \textbf{Node count} & \textbf{Connections count} & \textbf{Mean~\hedge cardinality} & \textbf{Target constraints} \\
            \hline
            \SetCell[r=8]{c} \rotatebox{90}{layered / feedforward}
            & 16\kk\_model & 20\kk & 766\kk & 37.3 & small \\
            & 64\kk\_model & 110\kk & 23\MM & 210.3 & small \\
            & 256\kk\_model & 216\kk & 90\MM & 417.2 & large \\
            & 1\MM\_model & 302\kk & 256\MM & 848.1 & large \\
            \hline
            & LeNet & 14\kk & 875\kk & 63.2 & small \\
            & AlexNet & 208\kk & 145\MM & 696.2 & large \\
            & VGG11 & 194\kk & 133\MM & 688.3 & large \\
            & MobileNet~V1 & 6.9\MM & 577\MM & 83.5 & large \\
            \hline
            \SetCell[r=4]{c} \rotatebox{90}{cyclic}
            & Allen V1 & 231\kk & 70\MM & 304.7 & large \\
            & 16\kk\_rand & $2^{14}$ & 2.1\MM & 128 & small \\
            & 64\kk\_rand & $2^{16}$ & 12.6\MM & 192 & small \\
            & 256\kk\_rand & $2^{18}$ & 67.4\MM & 256 & small \\
            \hline
        \end{tblr}
    }
    \caption{Spiking neural networks used in the experiments.}
    \vspace{-4pt}
    \label{tab:snns}
\end{table}

\begin{table}[t]
    \centering
    \resizebox{\columnwidth}{!}{
        \begin{tblr}{colspec={|c|c|c|c|}, width=1.1\columnwidth, rows={0.5em}, rowsep = 1pt} 
            \hline
            \SetCell[r=7]{c} \rotatebox{90}{\textbf{Algorithm}} & \textbf{Partitioning} & \textbf{Initial Placement} & \textbf{Placement Refinement} \\
            \cline{2-4}
            & \SetCell[r=2]{c} hierarchical (\ref{subsubsec:hierarchical_partitioning}) & \SetCell[r=2]{c} Hilbert curve (\ref{subsubsec:hilbert_initial_placement}) & \SetCell[r=4]{c} force-directed (\ref{subsubsec:force_directed_placement}) \\
            & & & \\
            \cline{2-4}
            & \SetCell[r=2]{c} hyperedge overlap (\ref{subsubsec:hyperedge_overlap_partitioning}) & \SetCell[r=2]{c} spectral (\ref{subsubsec:spectral_initial_placement}) & \\
            \cline{2-4}
            & & & \\
            \cline{2-4}
            & \SetCell[r=2]{c} sequential (\ref{subsubsec:sequential_partitioning}) & \SetCell[c=2,r=2]{c} minimum distance (\ref{subsubsec:minimum_distance_placement}) & \\
            & & & \\
            \hline
        \end{tblr}
    }
    \caption{Algorithms forming the compared mapping techniques.}
    \label{tab:algorithms}
\end{table}

\begin{figure}[t]
    \centering
    \vspace{-6pt}
    \includegraphics[width=\columnwidth]{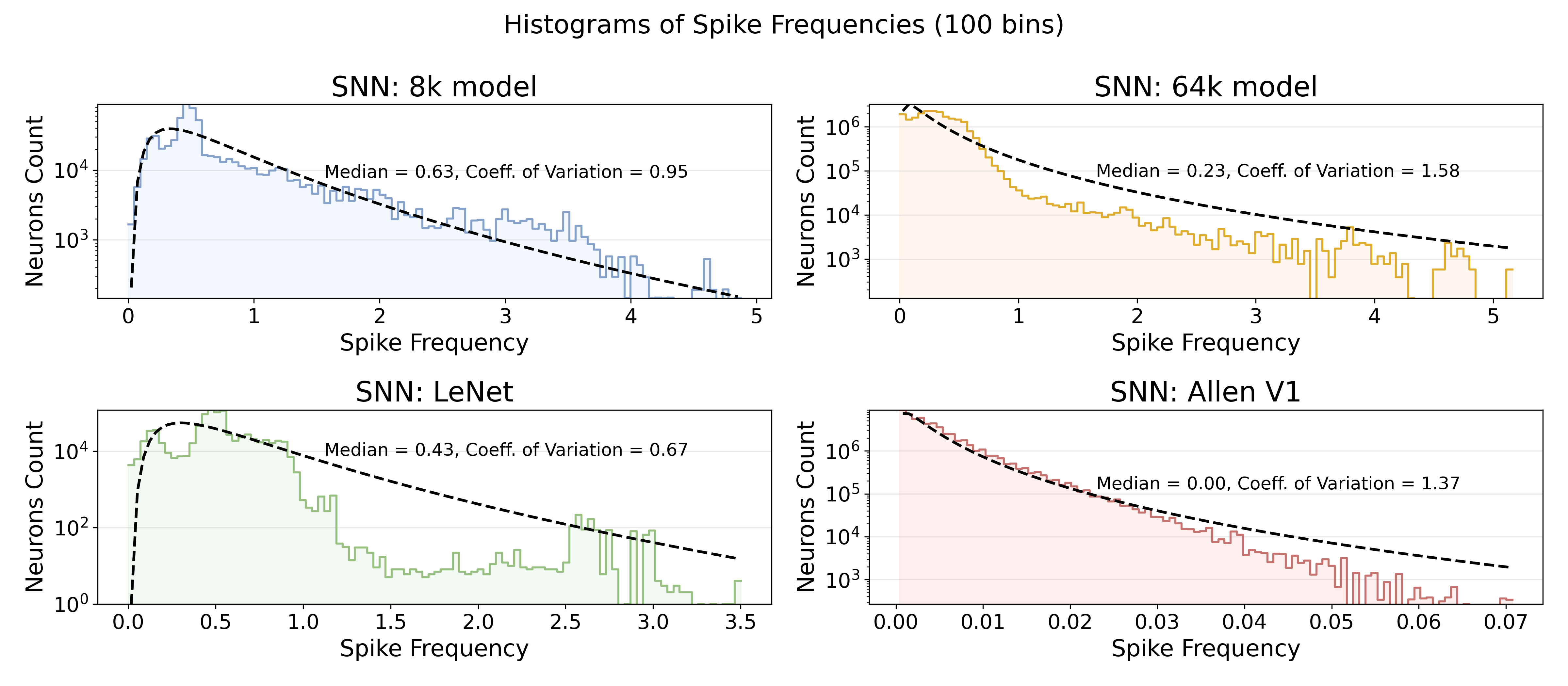}
    \vspace{-14pt}
    \caption{Spike frequencies distributions for four selected SNNs, fitted by a \mbox{log-normal} probability density function.}
    \vspace{-6pt}
    \label{fig:spike_frequencies_distributions}
\end{figure}

\begin{figure}[t]
    \centering
    \vspace{-4pt}
    \includegraphics[width=\columnwidth]{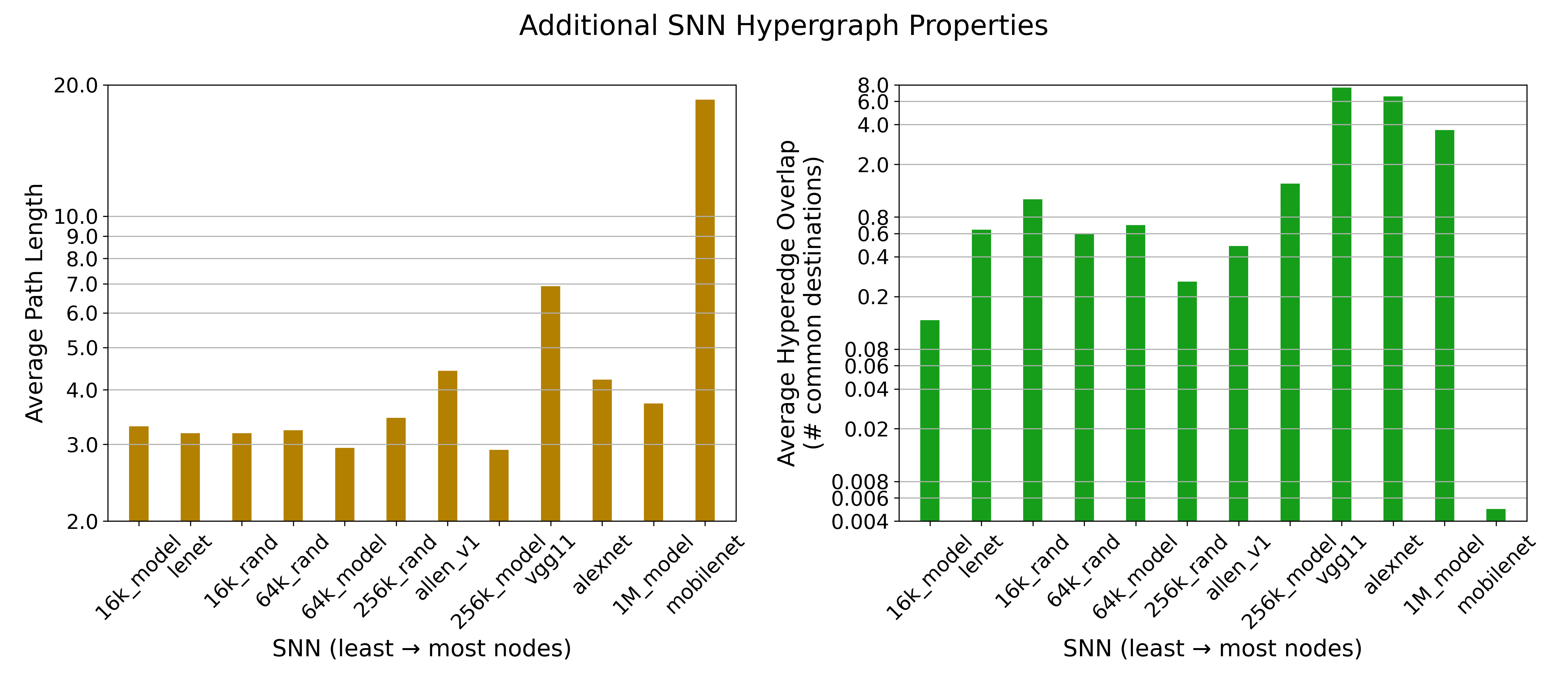}
    \vspace{-14pt}
    \caption{Average path length and \hedge overlap of considered SNNs.}
    \label{fig:snn_path_length_and_overlap}
\end{figure}

The algorithms we considered are summarized in Table~\ref{sec:algorithms}.
Our comparison spans all possible combinations of such partitioning, initial placement, and placement refinement techniques.
Our main baseline is formed by sequential partitioning and Hilbert curve placement with force-directed refinement.
For partitioning, however, we also compare against EdgeMap \cite{EdgeMap} and include the original unordered sequential partitioning from \cite{MappingVeryLargeSNNtoNHW}, which solely relies on the intrinsic order of nodes in the network, if any. 
Target NMH parameters have been introduced in Tab. \ref{tab:hw_costs}, we will rely on the "small" configuration up to $2^{26}$ connections, then switch to the "large" one.
This was required as bigger models readily exceed 4096 inbound connections per neuron when \hedge cardinality grows in the hundreds.
Relevant quality metrics have been defined in Tab.~\ref{tab:costs}; additionally, we will use the Energy-Latency Product (ELP) as a compound indicator of mapping performance.
While these metrics are not representative of any specific hardware platform, they have already been effectively used to rank mappings \cite{MappingVeryLargeSNNtoNHW, EdgeMap, HierarchicalSplitMapping}.
As anticipated, existing tools that implement heuristics from Sec.~\ref{subsubsec:hierarchical_partitioning} lack support for the constraints imposed by NMH.
At the same time, recent SNN mapping tools \cite{MappingVeryLargeSNNtoNHW, EdgeMap} don't have available artifacts.
Therefore, we re-implemented all presented heuristics, exploiting our \hgraph-based model's advantages whenever possible and ensuring compliance with NMH constraints.
This was done in Python 3.13 and with no explicit use of parallelism.
Experiments were carried out on an AMD EPYC 7453 @ 2.75GHz with 256GB of RAM.

The SNNs involved in our experiments are reported in Table~\ref{tab:snns}.
For layered SNNs, we considered eight convolutional neural networks.
They include four custom-built ones labeled "\texttt{$x$\_model}", with $x$ being the number of parameters.
Their architecture stacks VGG-like \cite{VGG} blocks until the desired number of parameters is reached, followed by global average pooling and a dense layer.
Then, we have four networks from literature: LeNet, AlexNet, and VGG11 trained on the Cifar10 dataset, and MobileNet for ImageNet 2012, all as implemented in \cite{keras, pytorch, VGG}.
We used SNNToolBox \cite{SNNtoolbox} to generate SNN versions of all networks and measure their spike frequencies while running inferences on 10\% of their respective datasets.

\begin{figure*}[th]
    \centering
    \vspace{-4pt}
    \includegraphics[width=\textwidth]{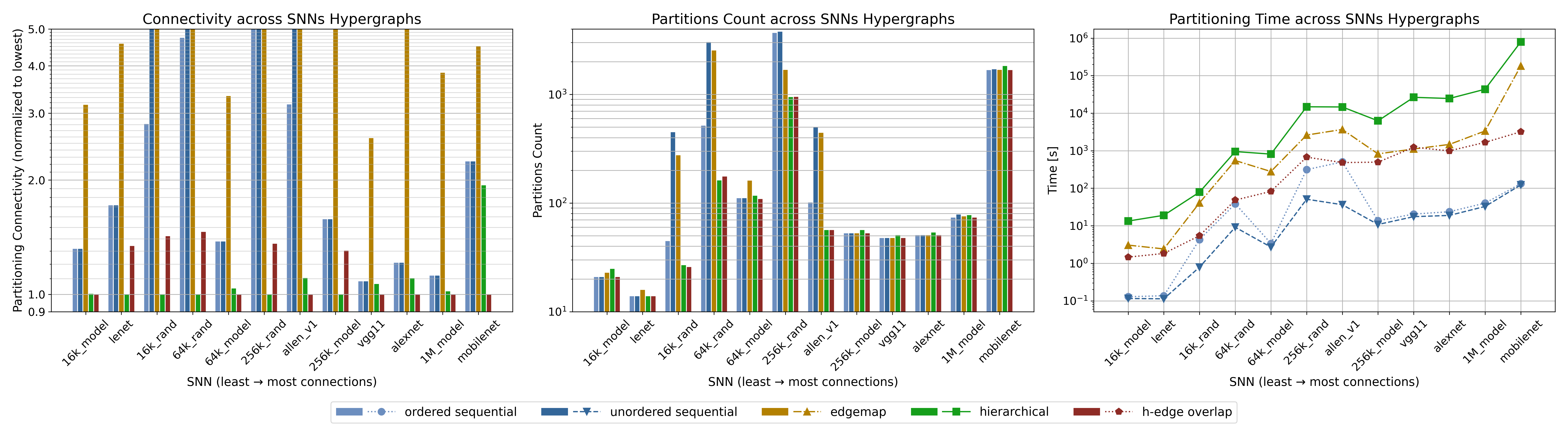}
    \vspace{-12pt} 
    \caption{Connectivity and execution time comparison between the considered partitioning heuristics.}
    \vspace{-14pt}
    \label{fig:partitioning_results}
\end{figure*}

As representatives of SNNs with arbitrary connectivity, we instead considered four biologically-inspired and cyclic networks.
First, we have the Allen V1 \cite{AllenV1}, a SNN modeled after a mouse's primary visual cortex.
Continuing, we took inspiration from liquid state machines \cite{LiquidStateMachine} and feedback SNNs \cite{FeedbackSNNs} to randomly generate three \hgraphs based on their topology.
These are named "\texttt{$x$\_rand}", $x$ being the number of nodes.
To construct them, we instantiate the required amount of nodes and randomly assign every one to a pair of coordinates inside the unit square.
Then, for each node, we sample its prospective connections count from a Poisson distribution of expected value equal to a fixed mean \hedge cardinality.
The target neurons of each \hedge follow a distance-dependent probability, where the likelihood of connection decays exponentially with the Euclidean distance over the previously assigned coordinates.
Finally, spike frequencies are sampled from a log-normal distribution with median 0.23 and coefficient of variation 1.58.
This choice is supported by biological evidence \cite{LognormInNeurons}, but also matches the spike frequency distribution of our other models, as shown in Fig.~\ref{fig:spike_frequencies_distributions}.
The resulting dense topology of strong connections is meant as a spike in difficulty -- pun intended -- for mapping.

Fig.~\ref{fig:snn_path_length_and_overlap} shows the average path length between any two nodes on the considered SNNs.
Such slow-growing values highlight the fact that both layered and cyclic SNNs are small-world networks.
It follows that, as the figure also shows, any pair of \hedges tends to overlap quite often, providing the basis for the exploitation of synaptic reuse.
As a control, our randomly-generated \hgraphs naturally fit among others for both measures.
MobileNet, instead, is a slight outlier, mainly due to its original ANN's high layer count, and shall serve as a test to see how well algorithms behave when reuse opportunities are scarce.






\subsection{Results Discussion}\label{subsec:results_discussion}

\begin{figure*}[th]
    \centering
    \vspace{-4pt}
    \includegraphics[width=\textwidth]{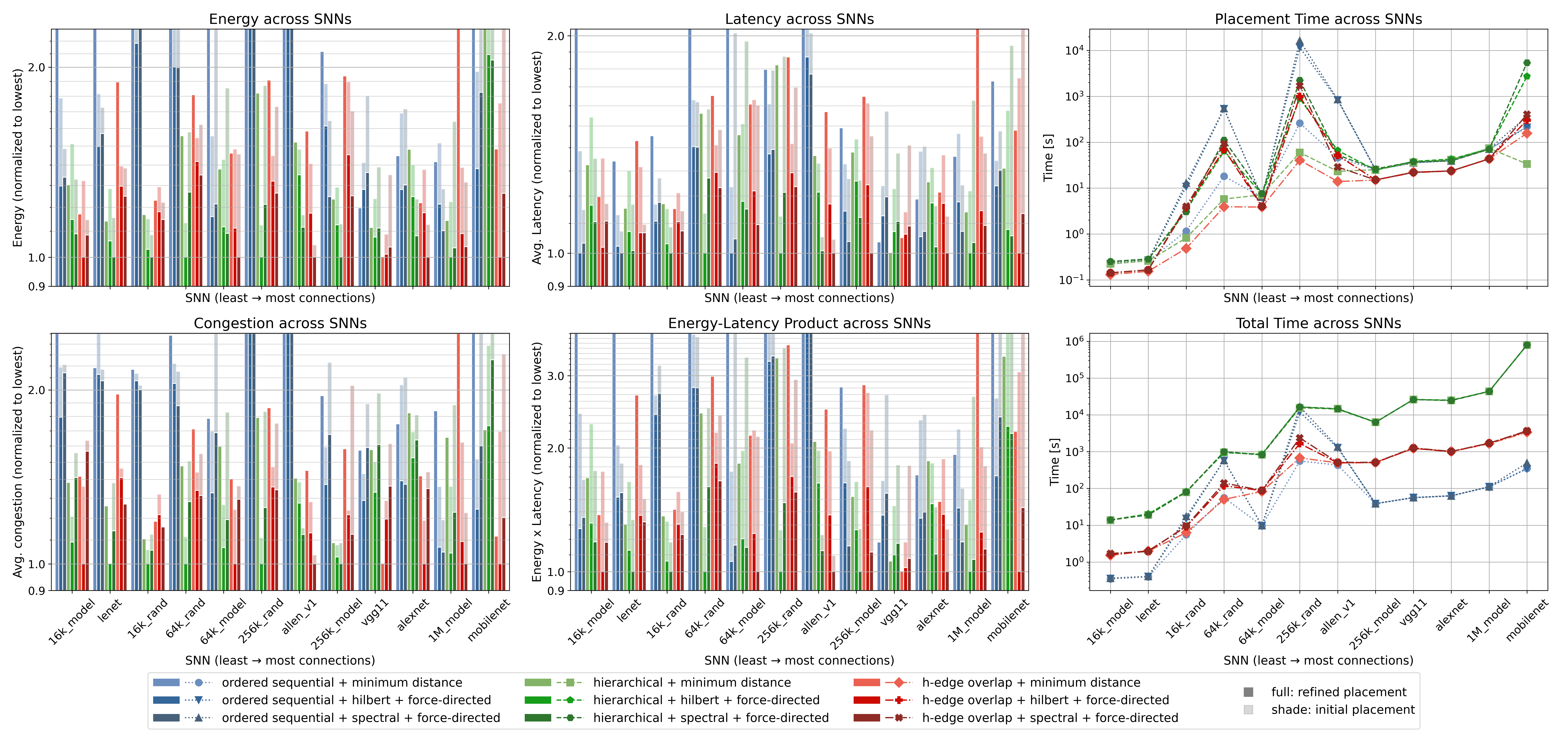}
    \vspace{-14pt}
    \caption{Mapping performance and construction time comparison between all placement techniques and partitioning algorithm pairs.}
    \vspace{-14pt}
    \label{fig:all_mapping_results}
\end{figure*}

In Fig.~\ref{fig:partitioning_results}, we report results for partitioning algorithms, compared by the achieved connectivity (Eq.~\ref{eq:connectivity}) and number of partitions.
Then, with Fig.~\ref{fig:all_mapping_results}, we show our results for complete mappings, after performing both partitioning and placement.

\subsubsection{Partitioning Results}\label{subsubsec:partitioning_results}


Looking at execution time first, we can clearly see three trends matching the presented algorithmic complexities.
All algorithms grow linearly with the count of nodes or \hedges.
At the bottom, unordered sequential partitioning is the fastest, having no further dependency.
Forming the middle line, with a linear dependency on \hedge cardinality, we have partitioning by \hedge overlap, EdgeMap, and ordered sequential (on cyclic \hgraphs).
All those algorithms iterate over each node's connections, or each \hedge's connectees, to rely upon guidance from either first- or second-order affinity.
The upper trend is constituted by hierarchical partitioning, with a quadratic dependency on \hedge cardinality.
In contrast to the growing number of neurons dictating the trend in execution time, the \hedge cardinality is relatively small and poses little limits to scalability, averaging at 316.4 on our SNNs.
Yet, being a multiplicative factor, it can be decisive in slowing down execution from hours to days.
We shall now evaluate whether each additional computational resources investment pays off in connectivity.

Hierarchical partitioning dominates for small \hgraphs, but the load of edge-coarsening soon hinders its execution time.
On average, its connectivity cost is $0.47\times$ that of sequential partitioning, and $0.95\times$ that of our overlap method.

Sequential partitioning, while simple, performs well on layered networks.
In spite of its lack of any active form of guidance, at times it reaches within just $1.02\times$ higher connectivity compared to the hierarchical method.
Even then, it is exceedingly fast and scalable, albeit so long as nodes are already ordered.
When arbitrarily connected networks are considered, however, it has to rely on our greedy ordering scheme to retain its quality of results.
This increases its execution time proportionally to \hedge cardinality, making it run in the same time as our overlap-based method, which nevertheless finds partitionings with at least $0.51\times$ the cost.
Foregoing such additional ordering step, the unordered sequential partitioning variant remains fast, but inevitably incurs up to $11.4\times$ its counterpart's connectivity.

Our new \hedge overlap-driven approach fits precisely between the slow hierarchical and fast sequential techniques under all metrics.
Its time spent on evaluating second-order affinity pays off, as it always achieves a connectivity within $0.52\times$ and $1.46\times$ that of hierarchical partitioning and $0.32\times$ to $0.91\times$ that of sequential.
On occasions, it even manages to produce up to $5\%$ fewer partitions than other methods, thanks to better exploitation of synaptic reuse.
Its execution time scales well on large networks, being within the expected complexity factor of \hedges cardinality -- at most $28.2\times$ -- slower than our baseline.
Moreover, on cyclic networks, our algorithm outspeeds all others but unordered partitioning; its results always reaching within $1.28\times$ of the best.
With both hierarchical partitioning and our overlap method leaning heavily on second-order affinity, we can trace back their small gap in results to a major difference.
The first repeatedly evaluates the affinity between each pair of nodes, as it builds all partitions simultaneously, thus managing to hide connections more uniformly.
The second measures affinity only locally to the present partition, as it fills them sequentially.
Even so, it still benefits from second-order affinity in ordering visits to nodes and \hedges, in turn only needing to scan each connection once.

The node-centric, graph-based scheme of EdgeMap, which relies foremost on source-destination connection strength, serves as our control experiment.
Iterating over each connection of each node, its execution time is comparable to our approach, however, its achieved connectivity is $8.5\times$ worse than ours, on average.
This highlights how second-order affinity results in significantly better guidance compared to individual connections, for a similar computational cost.

\subsubsection{Placement Results}\label{subsubsec:placement_results}

The differences in connectivity achieved by partitioning algorithms are largely reflected in their final mapping metrics, regardless of the placement algorithm used.
The choice of partitioning algorithm also doesn't seem to slow down placement, aside from minor variations attributable to slight gaps in partitions count.
So, let us consider for now the best mapping obtained for each partitioning scheme.
On average, hierarchical partitioning produces mappings with ELP $0.98\times$ relative to our method.
In turn, our overlap partitioning bests sequential with a mean $0.63\times$ ELP.
Across individual metrics, our method is always within $1.35\times$ of the best while consistently reaching values between $0.09\times$ and $1.01\times$ of any faster method.
In particular, for the Allen V1, our overlap partitioning plus refined spectral placement mapping technique unilaterally finds the best mappings in the least time compared to all other solutions.
This result underlines the key role that our identified affinities plays in handling the challenge of realistic SNNs.
Then again, on MobileNet, the largest network we considered, overlap-based partitioning also leads in mapping quality, solidifying its consistency at scale.

For initial placements, the Hilbert space-filling curve performs well, especially with few partitions or high path length networks (e.g. VGG11, MobileNet), where even far-apart cores are not excessively penalized.
However, its full reliance on nodes ordering starts to show limits on densely connected and cyclic SNNs.
Our proposed spectral layout is instead agnostic w.r.t. the number of cores, as it is driven directly by the affinities between nodes.
Consequently, after refinement, the spectral method gives, on average, $0.96\times$ the ELP, improving to $0.91\times$ across our four shortest path length networks.
On the \texttt{256\kk\_model}, where the path length is at its shortest, spectral further dominates with a mean $0.73\times$ energy and latency relative to Hilbert across all partitionings.
Nonetheless, looking at the interconnect's congestion shows that the Hilbert curve's mean resulting traffic is $0.92\times$ the spectral method's.
We ascribe this to communication being more evenly distributed over core-to-core links as a side-effect of the curve's locality-preserving properties.
That is in contrast to the spectral layout's many connections passing through the lattice's center following the spectrum's normalization during construction.
Both algorithms require negligible execution time, and neither has a significant impact on the execution time of the subsequent refinement.

\begin{figure*}[t]
    \centering
    \vspace{-4pt}
    \includegraphics[width=\textwidth]{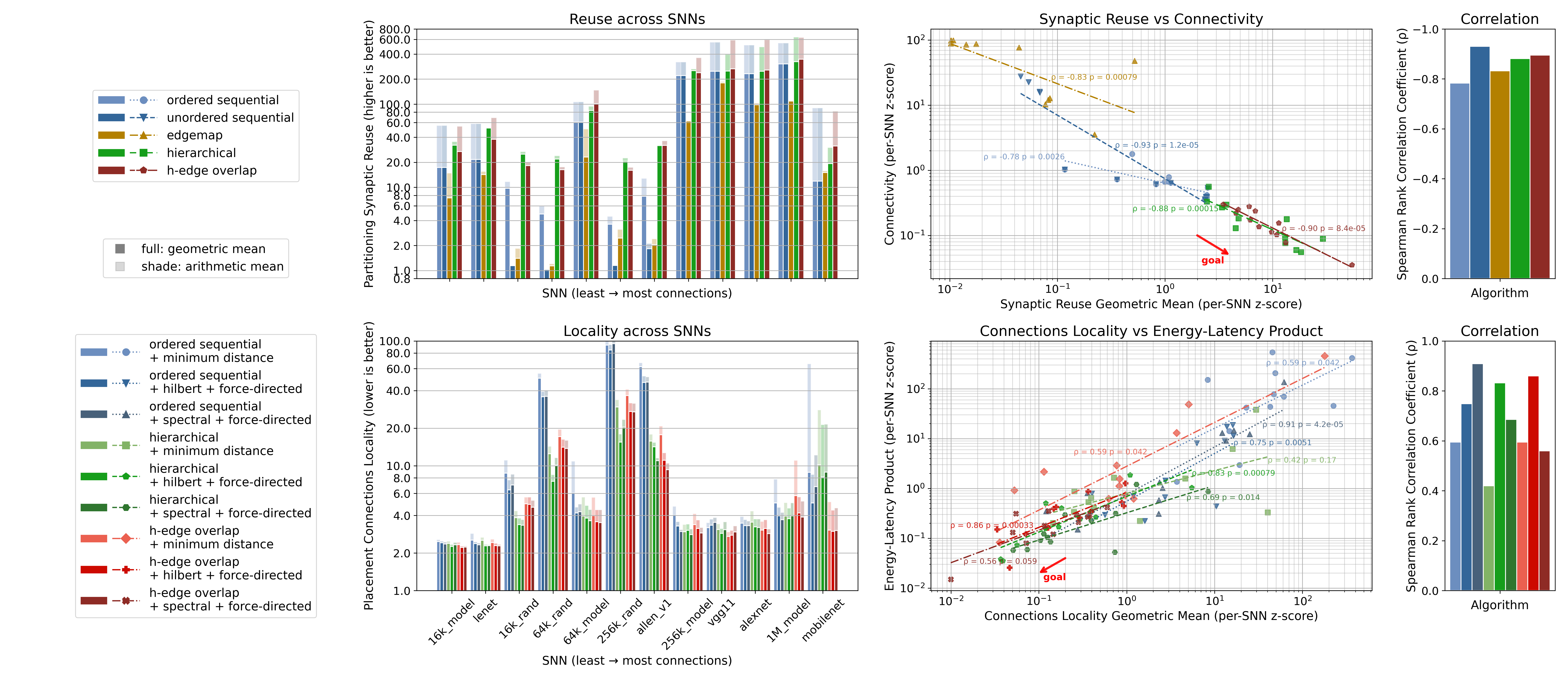}
    \vspace{-16pt}
    \caption{Synaptic reuse and connections locality measures and their correlation with mapping connectivity and energy-latency product.}
    \vspace{-16pt}
    \label{fig:concepts_results}
\end{figure*}

Comparing the refined layouts against their initial versions from Hilbert and spectral placement, we see that force-directed refinement consistently lowers every metric to $0.51$-$0.87\times$ its initial value.
Execution time for refinement, in general, grows linearly with the number of partitions that are being placed.
For most of our experiments, it usually is less than two minutes, as we tend to have $<$100 partitions, thus making partitioning time dominate.
Yet, when networks are dense enough for partitions to reach in the hundreds, mainly on our small NMH configuration, refinement starts occupying a significant part of the mapping process.
Surely, there is always the option to interrupt refinement early if, by then, results are deemed satisfactory, but we leave that up to the final user.
On the other hand, minimum distance placement starts to emerge as a faster alternative.
Its performance metrics are always within $2.18\times$ of the best achieved by force-directed refinement and $0.73$-$1.46\times$ of both initial placements.
However, it consistently runs in less than 2\kk~seconds.
Ultimately, force-directed refinement is the most flexible option, especially if a very good initial placement is available, but greedier alternatives will be faster at scale.

We thus conclude that relying on hypergraphs, their properties, and node affinities, be it with hierarchical partitioning, \hedge overlap, or spectral placement, yields very promising results, especially as SNN sizes grow and topologies vary.
We find that the regime of linear complexity in the SNN's connections, where affinities remain observable and actionable, holds particular potential for high-quality partitioning at scale.
Accordingly, the hierarchical method dominates for smaller SNNs, while overlap-based partitioning scales more gracefully; however, specific choices depend on available time and desired mapping quality.
Placement has more margin due to its smaller problem size, and running an ensemble of different techniques on a time limit -- then selecting the best final mapping -- is practicable.



\subsection{Verifying Mapping Properties}\label{subsec:verifying_mapping_properties}

To empirically verify our intuitions from Sec.~\ref{subsec:mapping_properties}, we hereby define and measure two quantities representing synaptic reuse and connections locality.  
We quantify synaptic reuse as the mean total number of individual inbound connections (synapses) per partition over the number of distinct inbound \hedges (axons) (partly based on \cite{HierarchicalSplitMapping}):
\vspace{2pt}
\begin{equation}\label{eq:synaptic_reuse}
    SR(G_S, G_P, \rho) \Def \mean_{p \in P} \left ( \frac{\sum_{(s, D) \in E_S} \abs{\{ d \in D \sothat \rho(d) = p \}}}{\abs{\{ (s, D) \in E_S \sothat \exists d \in D, \rho(d) = p \}}} \right )\text{.}
\end{equation}
Continuing, we quantify connections locality as the average number of lattice points enclosed by the convex hull (denoted by $conv$) defined around the cores connected by each \hedge:
\begin{equation}\label{eq:connections_locality}
    CL(G_P, H, \gamma) \Def \mean_{e = (s, D) \in E_P} \abs{conv(\{\gamma(p) \sothat p \in \{s\} \cup D\}) \,\cap\, H} \!\: \text{.}
\end{equation}
Fig.~\ref{fig:concepts_results} reports their measures across our experiments.
Data are presented using both geometric and arithmetic means as $mean$.
For synaptic reuse, the arithmetic mean reflects aggregate overlap, while the geometric mean emphasizes consistency across partitions and heavily penalizes low-overlap partitions.
Similarly, for connections locality, the arithmetic mean captures the overall average spatial footprint of \hedges, while the geometric mean emphasizes the typical footprint and is less influenced by a few very large \hedges.

Qualitatively, the trend for synaptic reuse tracks the SNNs' average hyperedge overlap, while connections locality is inversely related to the average path length (refer to Fig.~\ref{fig:snn_path_length_and_overlap}).
Then, all per-SNN performance differences between techniques are closely reflected by the hereby measures.
More in detail, we find that synaptic reuse's geometric mean perfectly follows connectivity, while the arithmetic mean tends to diverge.
Looking at our overlap-based algorithm, it often exhibits a noticeably higher arithmetic mean, implying that while it can reuse more synapses, it doesn't do so evenly across partitions, leading to more costly cuts.
Instead, the geometric mean reflects hierarchical partitioning's dominant results, suggesting that a uniform exploitation of synaptic reuse is preferable, albeit tougher to achieve.
For connections locality, there is little deviation between arithmetic and geometric mean, indicating that the spread of \hedges is mostly homogeneous.
Moreover, both means clearly reflect final mapping energy, latency, and congestion.

To assess the relationship between reuse/locality and mapping connectivity/ELP, Fig.~\ref{fig:concepts_results} also plots them against each other.
We then analyze the monotonic association between them using the Spearman’s rank correlation, with the resulting coefficients shown in the figure.
Because different SNNs exhibit widely different quality and property values, we here standardized both metrics per \hgraph (z-score).

Our conjectures, forming the basis for the present study, were that exploiting synaptic reuse leads to good partitionings and that improving connections locality provides better placements.
Those are indeed confirmed by the evident correlation between synaptic reuse and connectivity, and between connections locality and mapping performance.
Across all partitioning techniques, Spearman’s rank correlation between reuse and connectivity is strongly negative ($\rho \approx -0.86$ on average) with small deviation, proving higher synaptic reuse values consistently correspond to better partitions.
For placement, all correlations are significantly positive ($\rho \approx 0.69$ on average), confirming that mappings achieving lower ELP also exhibit better connections locality.
These consistent monotonic relationships demonstrate that our measures in Eqs.~\ref{eq:synaptic_reuse} and \ref{eq:connections_locality} are reliable predictors of solution quality.

We can consequently reaffirm that uniformly pursuing synaptic reuse and connections locality is a sound approach to raise mapping quality.
Additionally, these concepts hold firmly as the network is scaled or its topology is diversified.

\section{Conclusion}\label{sec:conclusion}

Bringing together nodes that partake in the same hyperedges emerges as the key policy identified in this work, one that we have shown to correlate strongly with high-quality mappings of spiking neural networks on neuromorphic hardware.
This principle benefits both partitioning -- by harnessing synaptic reuse -- and placement -- by increasing connections locality.
Taken together, these findings firmly cement hypergraphs as a sound and relevant abstraction for SNNs throughout their mapping, capable of naturally and precisely capturing the replication of spikes and their spread across cores.

Thereby, we opened a new avenue for high-quality SNN mapping based on hypergraphs.
Our preliminary selection of algorithms already demonstrates the practical value of this direction, delivering mappings up to twice as efficient as state-of-the-art graph-driven heuristics and remaining effective across layered, recurrent, and biologically plausible SNNs.
Notable gains come from hyperedge overlap-based partitioning, which scales linearly with problem size, and spectral initial placement, prized for its versatility.
These results indicate that hypergraph information can be exploited further while keeping computation scalable -- an essential requirement for approaching networks with billions of neurons.

To foster the development and comparison of further hypergraph algorithms applied to SNNs mapping, we plan to release our benchmark hypergraphs and algorithm implementations as open source.
At last, our future research efforts will focus on refining these hypergraph-based mapping heuristics and on the multi-chip generalization of the mapping problem.
We trust that this line of work lays the foundation for the continued improvement of mapping tools, paving the way for brain-scale SNNs on neuromorphic hardware.

\bibliographystyle{IEEEtran}
\bibliography{biblio}

\end{document}